\documentclass[12pt]{article}
\usepackage[latin1,utf8]{inputenc}
\usepackage[T1]{fontenc}
\usepackage[a4paper,left=2.5cm,right=2.5cm,top=2cm,bottom=1.5cm,footnotesep=.75cm,headheight=13.6pt]{geometry}
\usepackage{setspace}
\usepackage{amsmath}
\usepackage{graphicx}
\usepackage[small]{caption}
\usepackage{subcaption}
\usepackage{authblk}
\usepackage{rotating}
\usepackage{float}
\usepackage{pdflscape}
\usepackage{newtxmath}
\usepackage{appendix}
\usepackage{xcolor}

\title{The empirical distribution of sequential LS factors in Multi-level Dynamic Factor Models}

\author[1]{\large Gian Pietro Bellocca}
\author[1]{\large Ignacio Garr\'on}
\author[2]{\large C. Vladimir Rodríguez-Caballero}
\author[1]{\large Esther Ruiz\thanks{Corresponding author. E-mail address: ortega@est-econ.uc3m.es (E. Ruiz)}}
\affil[1]{\small{Department of Statistics, Universidad Carlos III de Madrid}\\
\small{C/ Madrid 126, 28903 Getafe (Spain)}}
\affil[2]{\small{Department of Statistics, ITAM, Mexico.}}
\affil[2]{\small{Department of Economics, Duke University, USA.}}

\date{\today}

\begin{document}

\maketitle

\begin{abstract}
The research question we answer in this paper is whether the asymptotic distribution derived by Bai (2003) for Principal Components (PC) factors in dynamic factor models (DFMs) can approximate the empirical distribution of the sequential Least Squares (SLS) estimator of global and group-specific factors in multi-level dynamic factor models (ML-DFMs). Monte Carlo experiments confirm that under general forms of the idiosyncratic covariance matrix, the finite-sample distribution of SLS global and group-specific factors can be well approximated using the asymptotic distribution of PC factors. We also analyse the performance of alternative estimators of the asymptotic mean squared error (MSE) of the SLS factors and show that the MSE estimator that allows for idiosyncratic cross-sectional correlation and accounts for estimation uncertainty of factor loadings is best. 
\end{abstract}



\setcounter{page}{1}

\newpage

\doublespacing

\section{Introduction}
Hashem Pesaran’s research agenda has profoundly influenced the econometric analysis of large panels with cross-sectional dependence. His work on the Common Correlated Effects (CCE) estimator (Pesaran, 2006) and his subsequent research on multifactor and hierarchical structures (Pesaran and Tosetti, 2011; Chudik and Pesaran, 2011; Chudik, Pesaran, and Tosetti, 2011; Chudik and Pesaran, 2015a) and multifactor representations establishes a coherent framework to model dependence driven by latent common shocks operating at different aggregation levels; see Chudik and Pesaran (2015b) for a comprehensive literature review. These ideas are closely related to dynamic factor models (DFMs) and their multi-level extensions, which have become central tools in empirical macroeconomics and finance; see also Ando and Bai (2016, 2017) for the conexion between panel models and DFMs.

In many empirical applications, latent factors may exhibit a hierarchical structure with global factors that load on all variables within the system, and group-specific factors that load only on subsets of variables. Multi-level dynamic factor models (ML-DFMs) capture this structure by imposing blocks of zero restrictions on the factor-loading matrix. A popular approach proposed by Breitung and Eickmeier (2016) to estimating ML-DFMs is sequential least squares (SLS), which extracts factors one level at a time and has been widely adopted due to its computational simplicity and empirical performance; see, for example, Kim, Kim and Bak (2017) and Kim and Kwark (2024). Despite its extensive use, formal inference on SLS factor estimates remains underdeveloped. In particular, explicit expressions for the asymptotic distribution of SLS-estimated factors are not available, although such results are essential for uncertainty quantification of the factors in applications such as scenario analysis and stress testing; see González-Rivera, Rodríguez-Caballero and Ruiz (2024).

In this paper, we first investigate whether the asymptotic distribution derived by Bai (2003) for principal components (PC) factors extracted from a DFM without a hierarchical structure can approximate the finite sample empirical distribution of SLS global and group-specific factors. This analysis is based on Monte Carlo simulations tailored to situations often faced in empirical applications. We show that the empirical distribution of SLS factors can be well approximated by the asymptotic results for PC factors in Bai (2003). Second, we also analyse the performance of alternative estimators of the asymptotic MSE available in the literature. In particular, we consider the popular heteroscedastic robust (HR) estimator proposed by Bai and Ng (2006), which is based on the unrealistic assumption of lack of idiosyncratic cross-correlation, and the estimator proposed by Fresoli, Poncela, and Ruiz (2025), which allows for it. We show that the performance of this latter MSE estimator improves over the HR estimator in the presence of idiosyncratic cross-correlation, while both are similar when the idiosyncratic components are uncorrelated. Furthermore, regardless of which of the two estimators of the asymptotic MSE of the factors is implemented, we analyse the performance of the subsampling correction proposed by Maldonado and Ruiz (2021) to incorporate uncertainty in the estimation of loadings in finite samples. We show that accounting for this uncertainty is crucial for the asymptotic distribution to be an appropriate approximation of the finite-sample distribution of SLS factors.

The rest of the paper is organised as follows. Section \ref{section:LR} provides a concise review of the literature on hierarchical DFMs. Section \ref{section:PC} introduces the ML-DFM framework and the SLS factor estimator. Section \ref{section:MC} presents the Monte Carlo results. Section \ref{section:final} concludes.

\section{Related literature on ML-DFMs}
\label{section:LR}

Pesaran’s emphasis on distinguishing between weak and strong cross-sectional dependence and, consequently, separating global from local sources of dependence, has motivated an extensive body of research on DFMs; see Bailey, Kapetanios, and Pesaran (2012) for a detailed discussion of weak, strong, and semi-strong common factors, which accommodate very general forms of cross-sectional dependence. The classification of strong, semi-strong, and weak cross-sectional dependence aligns naturally with ML-DFMs. Strong dependence arises from pervasive common factors that affect all units; semi-strong dependence reflects shocks that operate at intermediate aggregation levels, such as sectors or regions; and weak dependence corresponds to idiosyncratic components or localised interactions. 

In the DFM's literature, it is common to observe that some latent factors are group-specific and load only on particular groups of variables, with loadings equal to zero for all others; see Rodríguez-Caballero (2021), who incorporates a similar block factor structure in a panel data model within the CCE framework. This block structure may represent economic, geographical, cultural, or other characteristics; see, for example, Karadimiropoulou and Leon-Ledesma (2013), Moench, Ng and Potter (2013), Beck, Hubrich and Marcellino (2016), Breitung and Eickmeier (2016), Kim, Kim and Back (2017), Delle Chiaie, Ferrara and Giannone (2022), Lissona and Ruiz (2025), and Rodríguez-Caballero and Ruiz (2026), for empirical applications. The block structure in the loading matrix implies that the eigenvalues of the sample covariance matrix do not exhibit a clear break; see Onatski (2013), Han (2021), and Freyaldenhoven (2022), among others. Consequently, eigenvalue-based factor extraction procedures, such as PCA, are not well-suited to identifying group-specific factors.

Alternatively, factors can be extracted from ML-DFMs by imposing appropriate blocks of zero restrictions on the factor-loading matrix. Several estimation procedures address ML-DFMs. Early on, Wang (2010) proposes an iterated PC estimator initialised with PC factors. Along similar lines, Dias, Pinheiro, and Rua (2013) estimate ML-DFMs by minimising the overall idiosyncratic MSE subject to block-specific zero restrictions on the loadings.

Alternative approaches explore ML-DFMs from different perspectives. In state-space frameworks, Banbura, Giannone, and Reichlin (2011) and Lissona and Ruiz (2025) implement estimation through the Expectation–Maximisation (EM) algorithm of Doz, Giannone, and Reichlin (2012), initialising the factors with PC estimates. Within a Bayesian paradigm, Moench, Ng, and Potter (2013), Bai and Wang (2015), and Castelnuovo, Tuzcuoglu, and Uzeda (2025) develop Bayesian estimators. More recently, Han (2021) proposes an adaptive group-lasso shrinkage approach to identify global and group-specific factor spaces, while Chen (2023) introduces circularly projected common factors to recover latent hierarchical structures.

In this paper, we focus on the SLS estimator proposed by Breitung and Eickmeier (2016), which is based on the original estimator proposed by Wang (2010). After obtaining initial factors using Canonical Correlation Analysis (CCA), the global and group-specific factors are extracted sequentially via an iterative least-squares algorithm.
\footnote{Breitung and Eickmeier (2016) and Rodríguez-Caballero and Caporin (2019) extend the original SLS procedure, initially designed for non-overlapping block structures, to settings with overlapping blocks, which pose additional challenges for sequential estimation.} Several authors analyse the finite sample properties of SLS point estimates of the factors; see, for example, Breitung and Eickmeier (2016), Choi \textit{et al}. (2018), and Ergemen and Rodríguez-Caballero (2023) in a nonstationary setting. However, it is often of interest to also obtain measures of the uncertainty of the estimated factors, for example, when they are used to construct stressed economic scenarios, as in González-Rivera, Rodríguez-Caballero and Ruiz (2024). As far as we are concerned, there are no explicit expressions for the asymptotic distribution of the factors extracted using SLS. Note that Choi \textit{et al}. (2018) discuss the asymptotic behaviour of factors estimated by SLS, while Andreou \textit{et al}. (2019) derive an asymptotic expansion that could, in principle, be used to obtain the corresponding limiting distribution. However, neither contribution provides explicit expressions for the asymptotic distribution of the estimated factors. The closest related result is due to Lu, Jin, and Su (2025), who derive the asymptotic distribution of a two-step least squares estimator with initial PC factors and loadings in a ML-DFM with overlapping factors, as the number of blocks goes to infinity.

\section{Factor extraction in DFMs and ML-DFMs}
\label{section:PC}

This section introduces the DFM and ML-DFM together with their corresponding PC and SLS factor extraction procedures. We then present the asymptotic distribution of PC factors extracted from a DFM and describe how to estimate the corresponding MSE while allowing for idiosyncratic cross-sectional correlation and accounting for finite-sample uncertainty in factor-loading estimation.

\subsection{The DFM and the ML-DFM}

Let $\mathbf{Y}_t$ be the $N$-dimensional vector of stationary variables observed at time $t=1,...,T$, and consider the following popular approximate DFM
\begin{equation}
\label{eq:DFM}
\mathbf{Y}_t=\mathbf{\Lambda F}_t + \boldsymbol{\varepsilon}_t,
\end{equation}
where $\mathbf{\Lambda}=(\lambda_{ij}), i=1,\cdots,N$, and $j=1,\cdots,r$, is the $N\times r$ matrix of factor loadings, $\mathbf{F}_t= \left(F_{1t},...,F_{rt} \right)^{\prime}$ is the $r \times 1$ vector of underlying latent factors and $\boldsymbol{\varepsilon}_t=\left(\varepsilon_{1t},...,\varepsilon_{Nt} \right)^{\prime}$ is the $N \times 1$ vector of idiosyncratic components, which are allowed to be weakly cross-sectionally and serially correlated. $\boldsymbol{\Sigma}_{\varepsilon}$ denotes the covariance matrix of $\boldsymbol{\varepsilon}_t$. The number of factors $r$ is assumed to be known and fixed, with $r<N$. See Bai (2003) for a detailed discussion of the assumptions underlying the asymptotic theory of PC factor extraction, and Barigozzi and Hallin (2026) for a recent discussion of fundamental issues in the theory and practice of factor models.

The DFM in (\ref{eq:DFM}) is not identified. For any $r \times r$ full rank matrix $\mathbf{H}$, we can define a new set of common factors $\mathbf{F}_t^*=\mathbf{H F}_t $ and loadings $\mathbf{\Lambda}^*=\mathbf{\Lambda H}^{-1}$, which can generate an observationally equivalent system $\mathbf{Y_t}$ as follows:
\begin{equation}
\label{eq:identification}
\mathbf{Y}_t= \mathbf{\Lambda H}^{-1} \mathbf{H F}_t + \boldsymbol{\varepsilon}_t = \mathbf{\Lambda}^* \mathbf{F}_t^*+ \boldsymbol{\varepsilon}_t.
\end{equation}
The matrix $\mathbf{H}$ has $r^2$ free parameters and, therefore, $r^2$ restrictions are needed to identify $\mathbf{F}$ and $\mathbf{\Lambda}$. 

In the context of PC factor extraction, it is standard to impose the normalisation $\frac{1}{T}\mathbf{F}'\mathbf{F} = \mathbf{I}_r$, where $\mathbf{I}_r$ denotes the $r \times r$ identity matrix, which corresponds to $\frac{r(r+1)}{2}$ identifying conditions. This normalisation fixes the scale of the factors by standardising their second moments and removes indeterminacy due to scalar rescaling. It also enforces orthogonality across factors, ensuring that each factor captures distinct information.

The additional $\frac{r(r-1)}{2}$ identifying conditions follow by assuming that $\boldsymbol{\Lambda}^{\prime} \boldsymbol{\Lambda}$ is diagonal, with distinct positive entries ordered in decreasing magnitude. This set of restrictions diagonalises the cross-sectional product of the factor loadings and resolves the rotational indeterminacy of the factor space. As a result, the estimated factors identify the true factors up to a sign change; see the discussion by Bai and Ng (2013) on factor identification of DFMs.

As mentioned earlier, a wide range of empirical applications show that some latent factors are group-specific, loading on variables within a given group in $\mathbf{Y}_t$, and having zero loadings on the remaining variables. In this context, PC factor extraction is known to be inadequate as it does not take into account the zero restrictions in the matrix of loadings. Consequently, in the following, we also consider ML-DFMs in which the unobservable factors fall into two categories. The first consists of pervasive global factors that load on all variables in $\mathbf{Y}_t$. The second category contains group-specific factors, which are non-pervasive and load only on variables within a given group.

Let $y_{s,it}$ be the observation at time $t$ of the variable $i=1,...,N_s$ in group $s=1,...,S$, which is given by
\begin{equation}
\label{eq:ML-DFM_1}
y_{s,it}=\boldsymbol{\lambda}_{s,i}^{(g)\;\prime} \mathbf{G}_t+\boldsymbol{\lambda}_{s,i}^{(l)\;\prime} \mathbf{L}_{s,t}+\boldsymbol{\varepsilon}_{s,it},
\end{equation}
where $\boldsymbol{\lambda}_{s,i}^{(g)}$ and $\boldsymbol{\lambda}_{s,i}^{(l)}$ are $r_g$ and $r_s$ dimensional vectors of factor loadings, showing how each variable $i$ belonging to group $s$ is affected by global factors, $\mathbf{G}_t$, and group-specific factors $\mathbf{L}_{s,t}$, respectively. The $r_g \times 1$ vector $\mathbf{G}_t=\left(g_{1,t},...,g_{r_g,t} \right)^{\prime}$ contains the $r_g$ unobservable global factors, while the $r_s \times 1$ vector $\mathbf{L}_{s,t}$$=\left(l_{1,s,t},...,l_{r_s,s,t} \right)^{\prime}$ consists of the $r_s$ group-specific factors. Finally, the idiosyncratic component, $\varepsilon_{s,it}$, is defined as in the DFM in (\ref{eq:DFM}). In the ML-DFM in (\ref{eq:ML-DFM_1}), the total number of observations in all regions is $N=\sum_{s=1}^S N_{s}$. We assume that the number of groups $S$ is known and fixed. In practice, $S$ is much smaller than $ N$, and therefore we assume that $S \ll N$. We assume that the numbers of global and of group-specific factors are known, with the latter potentially varying across groups. 
Note that the literature often fixes the number of global and group-specific factors \textit{a priori}, usually setting one global factor and one group-specific factor per group; see, for example, Breitung and Eickmeier (2016), Choi \textit{et al.} (2018), Dias, Pinheiro and Rua (2020), and Artemova, Blasques and Koopman (2023). Furthermore, empirical work on ML-DFMs typically treats the partition of variables into groups as exogenous, so practitioners take the group structure as given.

We can rewrite the model in (\ref{eq:ML-DFM_1}) for all S groups in a system form
as 
\begin{equation}
\label{eq:ML-DFM_2}
\begin{pmatrix}
\mathbf{Y}_{1 t} \\
\vdots \\
\mathbf{Y}_{S t}
\end{pmatrix}
=
\begin{pmatrix}
\mathbf{\Lambda}_1^{(g)} & \mathbf{\Lambda}_1^{(l)} & 0         & \cdots & 0 \\
\mathbf{\Lambda}_2^{(g)} & 0         & \mathbf{\Lambda}_2^{(l)} & \cdots & 0 \\
\vdots   & \vdots    & \vdots    & \ddots & \vdots \\
\mathbf{\Lambda}_S^{(g)} & 0         & 0         & \cdots & \mathbf{\Lambda}_S^{(l)}
\end{pmatrix}
\begin{pmatrix}
\mathbf{G}_t \\
\mathbf{L}_{1t} \\
\mathbf{L}_{2t} \\
\vdots \\
\mathbf{L}_{St}
\end{pmatrix}
+
\begin{pmatrix}
\boldsymbol{\varepsilon}_{1 t} \\
\vdots \\
\boldsymbol{\varepsilon}_{S t}
\end{pmatrix},
\end{equation}
where $\mathbf{Y}_{st}$ is the $N_s \times 1$ vector of variables within group $s$ and $\boldsymbol{\varepsilon}_{st}$ is its corresponding vector of idiosyncratic components. The matrices $\mathbf{\Lambda}_s^{(g)}$ and $\mathbf{\Lambda}_s^{(l)}$ are the stacked over $i$ versions of $\boldsymbol{\lambda}_{s, i}^{(g)}$ and $\boldsymbol{\lambda}_{s,i}^{(l)}$, respectively. Stacking further across groups we obtain the following compact representation of the ML-DFM
\begin{equation}
\label{eq:ML-DFM}
\mathbf{Y}_t=\mathbf{\Lambda}^{\dagger} \mathbf{F}_t^{\dagger}+\boldsymbol{\varepsilon}_t,
\end{equation}
where $\mathbf{F}_t^{\dagger}=\left(\mathbf{G}_t, \mathbf{L}_{1,t},...,\mathbf{L}_{S,t} \right)^{\prime}$ collects all global and group-specific factors, and $\mathbf{\Lambda}^{\dagger}=\left[\mathbf{\Lambda}^{(g)}, \mathbf{\Lambda}^{(l)}\right]$ denotes the corresponding loading matrix with $\mathbf{\Lambda}^{(g)}=\left(\mathbf{\Lambda}_1^{(g)\prime}, \ldots, \mathbf{\Lambda}_S^{(g)\prime}\right)^{\prime}$ and $\mathbf{\Lambda}^{(l)}=\operatorname{diag}\left(\mathbf{\Lambda}_1^{(l)}, \ldots, \mathbf{\Lambda}_S^{(l)}\right)$. Note that the ML-DFM in (\ref{eq:ML-DFM}) is the DFM in (\ref{eq:DFM}) with zero restrictions on the loadings.

Identification remains necessary in ML-DFMs because rotations of the factor space leave the model unchanged; see Wang (2010), Bai and Wang (2015) and Breitung and Eickmeier (2016) for discussions. 
Consider the ML-DFM expressed as in (\ref{eq:ML-DFM}) and note that there is an $r \times r$ full rank matrix $H$ such that we can define a new set of common factors $\mathbf{F}_t^*=\mathbf{H F}^{\dagger}_t $ and loadings $\mathbf{\Lambda}^*=\mathbf{\Lambda}^{\dagger} \mathbf{H}^{-1}$, which can generate observationally equivalent $\mathbf{Y}_t$ as follows:
\begin{equation}
\label{eq:identification}
\mathbf{Y}_t=\mathbf{\Lambda}^{\dagger} \mathbf{H}^{-1} \mathbf{H F}^{\dagger}_t + \boldsymbol{\varepsilon}_t = \mathbf{\Lambda}^* \mathbf{F}_t^*+\boldsymbol{\varepsilon}_t.
\end{equation}
The main difference with the identification scheme described above for the DFM without group-specific factors is that in the ML-DFM the rotated matrix of loadings $\mathbf{\Lambda}^*$ should satisfy the zero restrictions, implying that the number of free parameters of $\mathbf{H}$ is smaller than $r^2$. In fact the number of free parameters of $\mathbf{H}$ is $r_g^2+\sum_{s=1}^S r_s^2 + r_g \sum_{s=1}^S r_s$; see Wang (2010).\footnote{See Appendix A for examples of the number of identifying restrictions in ML-DFMs with different factor structures.} To uniquely identify (up to a sign transformation) the factors and loadings of the ML-DFM, we follow Wang (2010) and Breitung and Eickmeier (2016) and consider the following identification restrictions\footnote{Note that these restrictions are appropriate to use the asymptotic distribution of PC factors, which as explained above assume that $\mathbf{\mathbf{\Lambda}}^{*\prime} \mathbf{\Lambda}^*$ is diagonal and the factors are orthonormal.}
\begin{enumerate}
\item $\forall s\in(1,S), \frac{\mathbf{L}_s^{\prime} \mathbf{L}_s}{T}=\mathbf{I}_{r_s}$ and $\mathbf{\Lambda}_s^{(l)\prime} \mathbf{\Lambda}_s^{(l)}$ is diagonal, which equals $\sum_{s=1}^S r_s^2$ restrictions.
\item $\frac{\mathbf{G}^{\prime}\mathbf{G}}{T}=\mathbf{I}_{r_g}$ and $\mathbf{\Lambda}^{(g)\prime} \mathbf{\Lambda}^{(g)}$ is diagonal, which amounts to $r_g^2$ restrictions.
\item $\forall s\in(1,S), \frac{\mathbf{L}_{s}^{\prime} \mathbf{G}}{T}=\mathbf{0}_{r_s\times r_g}$, which equals to $r_g\sum_{s=1}^S r_s$ restrictions 
\end{enumerate}


The third orthogonality conditions ensure that global factors do not absorb group-level dynamics and that local factors do not capture global variation, thereby allowing separate identification of global and local components. However, group-specific factors associated with different groups of variables do not need to be orthogonal.\footnote{Assuming that group-specific factors of different groups are orthogonal implies an over-identified model structure; see Breitung and Eickmeier (2016).} 

\subsection{PC and SLS estimators}

Define $\mathbf{Y}=\left(\mathbf{Y}_1, \mathbf{Y}_2,...,\mathbf{Y}_T \right) ^{\prime}$ as the $T \times N$ matrix of observations and consider the DFM in (\ref{eq:DFM}). Using the identification restrictions, the estimated PC factors, $\widetilde{\mathbf{F}}$, are $\sqrt{T}$ times the eigenvectors corresponding to the $r$ largest eigenvalues of $\mathbf{YY}^{\prime}$ arranged in decreasing order. Note that, by construction, $\frac{\widetilde{\mathbf{F}}^{\prime} \widetilde{\mathbf{F}}}{T}=I_r$. Using this normalization, the PC loadings are given by $\widetilde{\mathbf{\Lambda}}^{\prime}=\frac{1}{T}\widetilde{\mathbf{F}}^{\prime}\mathbf{Y}$, which by definition are such that $\widetilde{\mathbf{\Lambda}}^{\prime}\widetilde{\mathbf{\Lambda}}$ is diagonal; see Bai and Ng (2008). It is important to note that the PC factors estimate the space spanned by the factors and not the factors themselves, unless the identification restrictions are satisfied by the true factors. In the latter case, the estimated factors and the true factors are spanned in the same space.

The ML-DFM could also be estimated by PC without accounting for the zero restrictions in the loading matrix. However, these restrictions imply that the eigenvalues of $\mathbf{YY}^{\prime}$ do not exhibit a clear break and, consequently, eigenvalue-based factor extraction procedures, such as PC, are not well-suited to identify group-specific factors. Alternatively, Breitung and Eickmeier (2016) propose the SLS estimator to estimate the factors and loadings of the ML-DFM in (\ref{eq:ML-DFM_2}), which proceeds as follows:\footnote{In this paper, we implement the SLS estimator using the FARS package of Bellocca \textit{et al}. (2025).}

\begin{enumerate}
\item Initialisation. For $s=1,\ldots,S$, estimate $r_g+r_s$ factors separately in each group using PC. The PC factors extracted from different groups share a common component corresponding to the global factors. To estimate this component, apply Canonical Correlation Analysis (CCA) to identify the $r_g$ linear combinations of the group-specific PC factors that maximise cross-group correlation. The linear combinations associated with the largest canonical correlations provide initial estimates of the global factors, denoted by $\hat{ \mathbf{G}}_t^{(0)}$.\footnote{Instead of estimating $\hat{\mathbf{G}}^{(0)}_t$ using CCA, Wang (2010) proposes using PC to extract $r_g$ global factors from the entire data set $\mathbf{Y}_t$. The final estimates of the factors are the same regardless of how the initial estimates of the global factors are obtained.} Regress $\mathbf{Y}_t$ on $\hat{\mathbf{G}}_t^{(0)}$ and compute the residuals $\mathbf{U}_t$. Within each group, apply PC to $\mathbf{U}_t$ to obtain initial estimates of the group-specific factors, $\hat{\mathbf{L}}_{s,t}^{(0)}$.\footnote{Note that the initial estimates of the global factors, $\hat{\mathbf{G}}^{(0)}$ are orthonormal by construction. Similarly, for each group, the initial estimates of the group-specific factors are orthonormal. However, in practice, depending on the particular algorithm used to implement PC, the factors can have slight numerical deviations from orthonormality.}

\item Loadings update. Given the initial factor estimates, $\hat{\mathbf{G}}_t^{(0)}$ and $\hat{\mathbf{L}}^{(0)}_s, s=1,...,S$, estimate the global and group-specific loadings by cross-sectional LS. Note that in order to impose the zero restrictions in the matrix of loadings, they should be estimated either by restricted LS in the entire system, $\mathbf{Y}_{st}$, or group by group from the following model,
\begin{equation}
\label{eq:DFM_group}
\mathbf{Y}_{st}=\mathbf{\Lambda}_s \hat{\mathbf{F}}_{st} + \boldsymbol{\varepsilon}^*_{st},
\end{equation}
where $\mathbf{\Lambda}_s=\left( \mathbf{\Lambda}^{(g)}_s, \mathbf{\Lambda}^{(l)}_s \right)$ is the $N_s \times (r_g+r_s)$ matrix of loadings of the variables in group $s$, $\mathbf{Y}_s$, and $\hat{\mathbf{F}}_{st}=\left( \hat{\mathbf{G}}_t, \hat{\mathbf{L}}_{st} \right)^{\prime}$ is the corresponding vector of estimated factors. Denote the corresponding loadings as $\hat{\mathbf{\Lambda}}_{s}^{(g,0)}$ and $\hat{\mathbf{\Lambda}}_{s}^{(l,0)}$ for $s=1,\ldots,S$.\footnote{The estimated loadings of the global factors are orthogonal, similarly, the loadings of the group-specific factors are orthogonal within each group.}

\item Factors update. Re-estimate the global and group-specific factors by time-series LS from model (\ref{eq:ML-DFM}), with the true loadings substituted by $\hat{\mathbf{\Lambda}}_{s}^{(g,0)}$ and $\hat{\mathbf{\Lambda}}_{s}^{(l,0)}$. Denote the resulting estimated factors as $\hat{\mathbf{G}}_t^{(1)}$ and $\hat{\mathbf{L}}_{st}^{(1)}$, for $s=1,...,S$.\footnote{Note that, because not all columns of the matrix of loadings $\hat{\mathbf{\Lambda}}^{\dagger}$ are orthogonal, the OLS estimated factors do not necessarily are orthonormal.}

\item Iteration. Repeat steps 2 and 3 until the Residual Sum of Squares (RSS) converges.
\end{enumerate}

After convergence, the SLS estimator delivers factor estimates $\hat{\mathbf{G}}_t$ and $\hat{\mathbf{L}}_{s,t}$, along with their corresponding loading estimates $\hat{\mathbf{\Lambda}}_s^{(g)}$ and $\hat{\mathbf{\Lambda}}_s^{(l)}$ for $s=1,\ldots,S$.\footnote{In the context of a DFM with lagged factors, Breitung and Demetrescu (2025) show that the sequence of RSSs is non-decreasing.} It should be noted that the SLS estimated global and group-specific factors are mutually orthogonal by construction, satisfying the identifying restrictions $\frac{\hat{\mathbf{L}}_{s}^{\prime} \hat{\mathbf{G}}}{T}=\mathbf{0}_{r_s\times r_g}$. However, these final estimates do not necessarily satisfy the other identification restrictions due to the factors estimated in step 3 not being orthonormal. Consequently, for the estimated and true factors to be defined in the same space, the final estimates of the factors and loadings are transformed to satisfy $\frac{\hat{\mathbf{G}}^{\prime}\hat{\mathbf{G}}}{T}=\mathbf{I}_{r_g}$ and $\mathbf{\hat{\Lambda}}^{(g)\prime} \hat{\mathbf{\Lambda}}^{(g)}$ diagonal, and $\frac{\hat{\mathbf{L}}_s^{\prime} \hat{\mathbf{L}}_s}{T}=\mathbf{I}_{r_s}$ and $\hat{\mathbf{\Lambda}}_s^{(l)\prime} \hat{\mathbf{\Lambda}}_s^{(l)}$ diagonal, for $s=1,...,S$. In this way, the SLS estimated and true factors are defined in the same space and, consequently, they can not only be directly compared, but also there is not need to using any rotation matrix when looking at the asymptotic distribution.

Breitung and Eickmeier (2016) highlight the close relationship between the SLS estimator and the sequential PC approach of Wang (2010). Initialising the algorithm with CCA places the procedure in the neighbourhood of the global minimum and typically reduces the number of iterations required for convergence. Initialisation based on PC or CCA yields the same final estimates, although PC initialisation may be more convenient in settings with complex block structures; see Breitung and Eickmeier (2016), and Bellocca \textit{et al.} (2025) for comparative evidence. Furthermore, the SLS estimator is equivalent to the QML estimator under the assumption of Gaussian, independent and identically distributed (\textit{iid}) idiosyncratic components when treating the common factors as unknown fixed parameters.

\subsection{Asymptotic distribution of PC factors}

Consider the DFM in (\ref{eq:DFM}). Stock and Watson (2002) show that, when the idiosyncratic cross-sectional correlations are weak, and the common factors are pervasive, the factor space spanned by the PC estimator, $\widetilde{\mathbf{F}}_t$, is consistently estimated as both $N$ and $T$ tend simultaneously to infinity. Bai (2003) extends this result by establishing consistency of PC factor estimates under serial and cross-sectional correlation and heteroscedasticity of the idiosyncratic components, again under joint asymptotics $N,T\to\infty$. Furthermore, provided that $\frac{\sqrt{N}}{T}\rightarrow 0$, Bai (2003) derives the following asymptotic distribution of PC factors 
\begin{equation}
\label{eq:asymptotic_1}
\sqrt{N}\left( \widetilde{\mathbf{F}}_{t}-\mathbf{F}_{t}\right)\stackrel{d} {\rightarrow} N \left( 0,\mathbf{\Sigma} _{\Lambda}^{-1}\mathbf{\Gamma} _{t}\mathbf{\Sigma} _{\Lambda}^{-1}\right),
\end{equation}
where $\frac{1}{N} \mathbf{\Lambda}'\mathbf{\Lambda} \to \mathbf{\Sigma}_{\Lambda}$ as $N \to \infty$, with $\mathbf{\Sigma}_{\Lambda}>0$ a positive definite $r\;\times\;r$ non-random matrix with distinct entries, and $\mathbf{\Gamma}_t = \lim\limits_{N \to \infty} {\frac{1}{N}\sum_{i=1}^N \sum_{j=1}^N \boldsymbol{\lambda}_i^{\prime} \boldsymbol{\lambda}_j E(\varepsilon_{it} \varepsilon_{jt})}$ which captures cross-sectional dependence and heteroscedasticity in the idiosyncratic components.

Several comments about the asymptotic distribution in (\ref{eq:asymptotic_1}) are relevant for this paper. First, note that the asymptotic distribution in (\ref{eq:asymptotic_1}) is stated without an explicit rotation matrix. This is because it is assumed that the true factors and loadings satisfy the identifying restrictions and, consequently, the true and estimated factors span the same space. Under the normalisation $T^{-1}\mathbf{F}'\mathbf{F}=\mathbf{I}_r$, Bai and Ng (2013) show that the rotation matrix $\mathbf{H}$ usually appearing in the asymptotic distribution satisfies $\mathbf{H}=\mathbf{I}_r+\mathbf{O}_p(\alpha_{NT}^{-2})$, where $\alpha_{NT}=\min(\sqrt N,\sqrt T)$, and therefore converges in probability to the identity matrix. Jiang, Uematsu and Yamagata (2023) refer to this specification as an \emph{identifiable pseudo-true model} and analyse the asymptotic distribution of rotated factors when identification is imposed through a deterministic rotation. Second, Bai and Ng (2006) further show that, when the idiosyncratic components are serially uncorrelated, the limiting distributions in (\ref{eq:asymptotic_1}) are asymptotically independent across $t$. Third, asymptotic normality holds without assuming Gaussianity of neither the common factors nor the idiosyncratic components. Bai (2003) notes that the restriction $\frac{\sqrt{N}}{T}\to 0$ is mild; therefore, asymptotic normality is the more prevalent situation in empirical applications. Fourth, at each time $t$, the asymptotic MSE of $\sqrt{N} \widetilde{\mathbf{F}}_t$ in (\ref{eq:asymptotic_1}) is diagonal only when the idiosyncratic errors are cross-sectionally uncorrelated and homoscedastic, that is, $E\left(\varepsilon_{it} \varepsilon_{jt}\right)=0,\;\forall\;i\neq j$, and $E\left(\varepsilon_{it}^2\right)=\sigma^2_{\varepsilon},\;\forall\;i$. Under these conditions, $\mathbf{\Gamma}_t=\mathbf{\Sigma}_{\Lambda}$, and the asymptotic MSE reduces to $\sigma^2_{\varepsilon} \mathbf{\Sigma}_{\Lambda}^{-1}$, which is diagonal under the identifying restriction that $\mathbf{\Lambda}^{\prime} \mathbf{\Lambda}$ is diagonal. However, when the idiosyncratic components exhibit cross-sectional correlation and/or heteroscedasticity, the estimated PC factors are correlated across different realizations and the asymptotic MSE is no longer diagonal. In this case, the off-diagonal elements of the MSE matrix depend in a non-trivial way on the magnitudes of the idiosyncratic variances, as well as on the signs and magnitudes of the factor loadings and the idiosyncratic covariances.

In practice, inference on PC factors based on the asymptotic distribution in (\ref{eq:asymptotic_1}) requires an estimate of the asymptotic MSE of $\widetilde{F}_t$, which can be estimated in finite samples as follows:
\begin{equation}
\label{eq:avar}
\widehat{Avar}(\widetilde{\mathbf{F}}_t)=\frac{1}{N}\left( \frac{\tilde{\mathbf{\Lambda}}^{\prime} \tilde{\mathbf{\Lambda}}}{N}\right) ^{-1} \widetilde{\mathbf{\Gamma}}_t \left( \frac{\tilde{\mathbf{\Lambda}}^{\prime} \tilde{\mathbf{\Lambda}}}{N}\right) ^{-1},
\end{equation}
where $\widetilde{\mathbf{\Gamma}}_t$ denotes an estimator of $\mathbf{\Gamma}_t$. Then, $100(1-\alpha)\%$ confidence regions for the factors are given by
\begin{equation}
\label{eq:asymregion}
R(\mathbf{F}_t)=\left\lbrace \mathbf{F}_t|\left(\mathbf{F}_t-\widetilde{\mathbf{F}}_t \right)^{\prime}\left[ \widehat{Avar}(\widetilde{\mathbf{F}}_t)\right] ^{-1}\left(\mathbf{F}_t-\widetilde{\mathbf{F}}_t \right) \leq \chi^2_{r(1-\alpha)}\right\rbrace,
\end{equation} 
where $\chi^2_{r(1-\alpha)}$ is the $(1-\alpha)$ quantile of the $\chi^2$ distribution with $r$ degrees of freedom.

Finally, with respect to the estimation of $\mathbf{\Gamma}_t$, it can be estimated by the following heteroscedasticity-robust (HR) estimator
\begin{equation}
\label{eq:gamma_1}
\widetilde{\mathbf{\Gamma}}^{HR}_{t}=\frac{1}{N} \widetilde{\mathbf{\Lambda}}^{\prime} \widetilde{\mathbf{\Sigma}}_{\varepsilon t}^* \widetilde{\mathbf{\Lambda}}=\frac{1}{N}\sum_{i=1}^{N}\tilde{\boldsymbol{\lambda}}_{i}\tilde{\boldsymbol{\lambda}}_{i}^{\prime }\tilde{\varepsilon}_{it}^{2},
\end{equation}
where $\widetilde{\mathbf{\Sigma}}_{\varepsilon t}^*$ is a diagonal matrix with $\widetilde{\varepsilon}^2_{it}$ on the main diagonal and $\tilde{\varepsilon}_{it}=y_{it}-\tilde{\boldsymbol{\lambda}}_{i}^{^{\prime}}\widetilde{\mathbf{F}_{t}}$ are the PC residuals that estimate the idiosyncratic components.

The HR estimator of $\mathbf{\Gamma}_t$ is only consistent in the absence of idiosyncratic cross-sectional correlation. Even though Bai and Ng (2006) propose using it even in the presence of weak idiosyncratic cross-sectional correlation, arguing that explicitly estimating small cross-correlations may increase sampling variability and lead to efficiency losses, when cross-sectional correlation is non-negligible, $\tilde{\mathbf{\Gamma}}_t^{HR}$ becomes inconsistent and the associated confidence regions in (\ref{eq:asymregion}) are no longer valid. In such cases, computing the asymptotic covariance matrix under the incorrect assumption that $\mathbf{\Sigma}_{\varepsilon}$ is diagonal may either overestimate or underestimate the true asymptotic covariance of the factor estimates.\footnote{The error in estimating the asymptotic covariance matrix of the factors that arises from incorrectly assuming that $\mathbf{\Sigma}_{\varepsilon}$ is diagonal depends on the matrix $\mathbf{A}=\mathbf{\Sigma}_{\varepsilon}-\mathbf{\Sigma}_{\varepsilon}^*$, where $\mathbf{\Sigma}_{\varepsilon}^*=\text{diag}(\mathbf{\Sigma}_{\varepsilon})$ denotes the diagonal matrix formed from the main diagonal of $\mathbf{\Sigma}_{\varepsilon}$. By construction, $\mathbf{A}$ has zero diagonal elements and potentially non-zero off-diagonal elements. As a result, $\mathbf{A}$ is a symmetric indefinite matrix, and the $r$ quadratic forms appearing in $\mathbf{\Lambda}^{\prime}\mathbf{A\Lambda}$ may take either positive or negative values, depending on the magnitudes and signs of the factor loadings and the idiosyncratic covariances.} Alternatively, as proposed by Fresoli, Poncela and Ruiz (2025), we also estimate $\mathbf{\Gamma}_t$ allowing for cross-sectional correlation as follows:
\begin{equation}
\label{eq:FPR}
\widetilde{\mathbf{\Gamma}}^{FPR}=\frac{1}{N} \widetilde{\mathbf{\Lambda}} \widetilde{\mathbf{\Sigma}}_{\varepsilon} \widetilde{\mathbf{\Lambda}},
\end{equation}
where the ij'th element of $\widetilde{\mathbf{\Sigma}}_{\varepsilon}$ is given by
\begin{equation}
\label{eq:Sigma_threshold}
\widetilde{\sigma}_{ij}= \frac{1}{T} \sum_{t=1}^T \tilde{\varepsilon}_{it} \tilde{\varepsilon}_{jt} I\left(\mid \hat{\sigma}_{ij}\mid \geq c_{ij} \right), 
  \end{equation}
with
\begin{equation}
\label{eq:Sigma_sample}
\hat{\sigma}_{ij}=\frac{1}{T} \sum_{t=1}^T \tilde{\varepsilon}_{it} \tilde{\varepsilon}_{jt}.
\end{equation}

In equation (\ref{eq:Sigma_threshold}), the threshold $c_{ij}$ is defined 
as follows
\begin{equation}
\label{eq:c}
c_{ij}=\delta \omega_{NT} \widehat{\theta} _{ij}^{\frac{1}{2}},
\end{equation}
where $\widehat{\theta}_{ij}=\widehat{Var}\left[\tilde{\varepsilon}_{it} \tilde{\varepsilon}_{jt} \right]= \frac{1}{T}\sum_{t=1}^T \left(\tilde{\varepsilon}_{it} \tilde{\varepsilon}_{jt} - \tilde{\sigma}_{ij}\right)^2$, $\delta$ is a threshold level\footnote{See Fresoli, Poncela and Ruiz (2025) for how to determine the value of $\delta$.}, and 
$\omega_{NT}=1/\sqrt{N}+\sqrt{\mathrm{log}(N)/T}$. Note that the estimator in (\ref{eq:Sigma_sample}) requires stationarity, and is the same for every $t$.

Neither of the two estimators of the MSEs of the PC factors discussed above accounts for the additional uncertainty introduced by estimating the factor loadings. To address this issue, we adopt the subsampling correction proposed by Maldonado and Ruiz (2021), which adjusts the finite-sample approximation of the asymptotic MSE in (\ref{eq:avar}) to account for loading-estimation uncertainty.

\subsection{Asymptotic properties of SLS factors}

Consider now the asymptotic properties of SLS factors. First, note that initialising the procedure with CCA yields consistent estimates of the global factors, which in turn lead to consistent estimates of the group-specific factors after projection. Therefore, SLS inherits consistency from the initial factor estimates when the number of factors is known; see also the consistency results of Breitung and Demetrescu (2025) for the SLS estimator in the context of a DFM with lagged factors.\footnote{Lin and Shin (2023) argue that the identification of the number of factors can be difficult. Consequently consistent estimation of global factors using CCA becomes challenging when two or more groups share group-specific factors, as they may be mistakenly identified as global factors.}

As mentioned above, when the number of groups remains fixed with $N$, ML-DFM can be seen as a DFM with $r$ factors. Wang (2010) argues that, when the initial estimates are obtained using PC, the asymptotic results in Bai (2003) extend to the multi-level setting, subject to modifications that account for the zero restrictions on the loading matrix. Under these conditions, the convergence rates of global and group-specific factors coincide.\footnote{Wang (2010) also derives results under asymptotics in which the number of blocks diverges, whereas Lu, Jin and Su (2025) derive the asymptotic distribution of a two-step PC estimator for ML-DFMs with overlapping blocks when the number of blocks tends to infinity, noting that their approach does not apply when one cross-sectional dimension is fixed.}

Because the SLS estimator converges to the same solution regardless of whether the algorithm is initialised using PC or CCA, the asymptotic distribution derived by Bai (2003) should also apply in the latter case. Choi \textit{et al.} (2018) argue that the asymptotic distribution of global factors estimated by SLS coincides with that of PC factors in the standard DFM. In addition, they show that the residuals obtained after removing the global component are consistent, suggesting that the same asymptotic distribution should govern the group-specific factors. However, they do not derive the explicit covariance matrix. Andreou \textit{et al.} (2019) provide an asymptotic expansion that can be used to obtain the limiting distributions of the factors and loadings estimated via CCA. 

Consequently, it could be argued that the asymptotic distribution of SLS factors is given by (\ref{eq:asymptotic_1}) with $\widetilde{\mathbf{F}}_t$ denoting the corresponding SLS estimates.\footnote{Recall that the asymptotic distribution in (\ref{eq:asymptotic_1}) lacks the usual rotation matrix because the identifying assumptions are assumed to be satisfied. Therefore, even though orthogonality of the group-specific factors among groups are not needed for identification of the ML-DFM, we are assuming them to avoid using a rotation matrix.} In finite samples, we approximate the asymptotic MSE using (\ref{eq:avar}), where  $\widetilde{\mathbf{\Lambda}}$ corresponds to the loading matrix with loadings replaced by their SLS estimates, and $\widetilde{\mathbf{\Gamma}}_{t}$ is estimated under the assumption of cross-sectionally uncorrelated idiosyncratic components, as in (\ref{eq:gamma_1}), or allowing for cross-sectional correlation, as in (\ref{eq:Sigma_threshold}). In all cases, the subsampling procedure provides a convenient way to incorporate the uncertainty associated with loading estimation.

\section{Monte Carlo simulations}
\label{section:MC}

In this section, we assess whether the asymptotic distribution in (\ref{eq:asymptotic_1}) derived by Bai (2003) for PC factors provides an accurate approximation to the finite-sample distribution of the factors extracted using the SLS procedure in ML-DFMs. We also explore the performance of alternative estimators of the asymptotic MSE in this setting.

We consider a Monte Carlo design based on the ML-DFM in (\ref{eq:ML-DFM_1}) with two blocks ($S = 2$), one global factor ($r_g = 1$), and one group-specific factor per block ($r_1 = r_2 = 1$).\footnote{Note that this particular structure is not relevant for the results about whether the asymptotic distribution of Bai (2003) is a good approximation of the empirical distribution of SLS factors. The number of groups, the number of global and group-specific factors, and the cross-sectional dimension within each group, affect the MSEs themselves but not whether they can be well approximated by using the MSE implied by the asymptotic distribution of Bai (2003).} The temporal dimensions considered are $T \in \{50, 100, 500\}$, and the cross-sectional dimensions per block are $N_1, N_2 \in \{25, 75, 300\}$. 

\begin{figure}[h!]
\begin{center}
\includegraphics[trim=15mm 15mm 0mm 18mm, clip, width=1\textwidth]{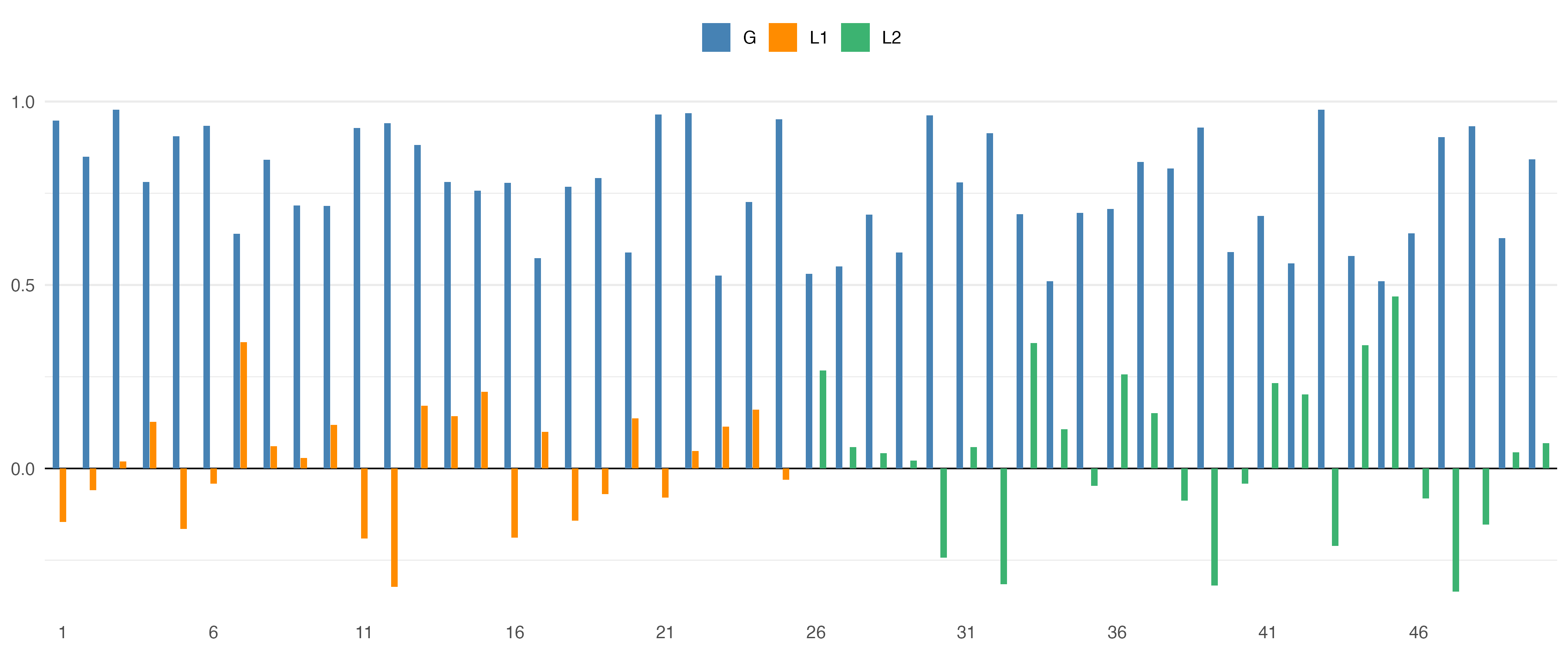}
\caption{Loadings of the global factor (blue), the first group-specific factor (orange), and the second group-specific factor (green) in the ML-DFM with $N_1 = N_2 = 25$. }
\label{fig:loadings_MLDFM}
\end{center}
\end{figure}

The loadings are independently drawn once from a $U(0.5,1)$ distribution and subsequently transformed to be orthogonal, resulting in the loading matrix $\mathbf{\Lambda}$, which is kept fixed in all replications. As an illustration, Figure \ref{fig:loadings_MLDFM} plots the loadings generated when $N_1 = N_2 = 25$. The loadings of the global factor are substantially higher than those of the group-specific factors. Further, the two non-pervasive factors show similar scales across both blocks. 

All three factors are generated once as independent AR(1) processes with zero mean, autoregressive parameter $\phi = 0.5$, and innovation variance $\sigma^2_{\eta} = 1 - \phi^2$, so that they have unit variance. After transforming the factors to be orthonormal, they are denoted $G_t$, $L_{1t}$, and $L_{2t}$. The factors are kept fixed in all Monte Carlo replications.\footnote{Note that the asymptotic distribution of PC-based factors is derived conditional on a fixed realization of the factors; see Barigozzi, Fresoli, and Ruiz (2026).}

For each replication $m=1,\ldots , M$, we generate the idiosyncratic components as Gaussian white noise with covariance matrix $\mathbf{\Sigma}_{\varepsilon}$, where $\sigma_i^2 = c\,u_i$ and $u_i = 1$ (homoscedasticity) or $u_i \sim U(0.5,2)$ (heteroscedasticity). The constant $c=0.25$ controls the signal-to-noise ratio.\footnote{Given that the factors are orthogonal, the relative magnitude between the absolute value of the loadings and the idiosyncratic covariances, in particular, $\mathbf{\Lambda}^{\prime} \mathbf{\Sigma}^{-1}_{\varepsilon} \mathbf{\Lambda}$, measures the strength of the factors. Obviously, for a given  $N$, the larger this strength is, the smaller is the MSE of the estimated factors. Once more, the particular values of the matrix $\mathbf{\Lambda}^{\prime} \mathbf{\Sigma}^{-1}_{\varepsilon} \mathbf{\Lambda}$ do not affect whether the empirical distribution of Bai (2003) is a good approximation of the empirical distribution of the SLS factors.} Idiosyncratic cross-sectional dependence is introduced through a Toeplitz structure as follows:
$$\sigma_{ij} = \sigma_i \sigma_j \tau^{|i-j|}, \qquad i = 1,\ldots,N,\; j = i+1,\ldots,N,$$
where $\tau \in \{0,-0.5\}$. After generating $\sigma_{ij}$, the columns are subsequently permuted. When $\tau \neq 0$, the idiosyncratic components exhibit cross-sectional correlation.

\begin{figure}[h!]
\begin{center}
\includegraphics[trim=15mm 15mm 0mm 18mm, clip, width=1\textwidth]{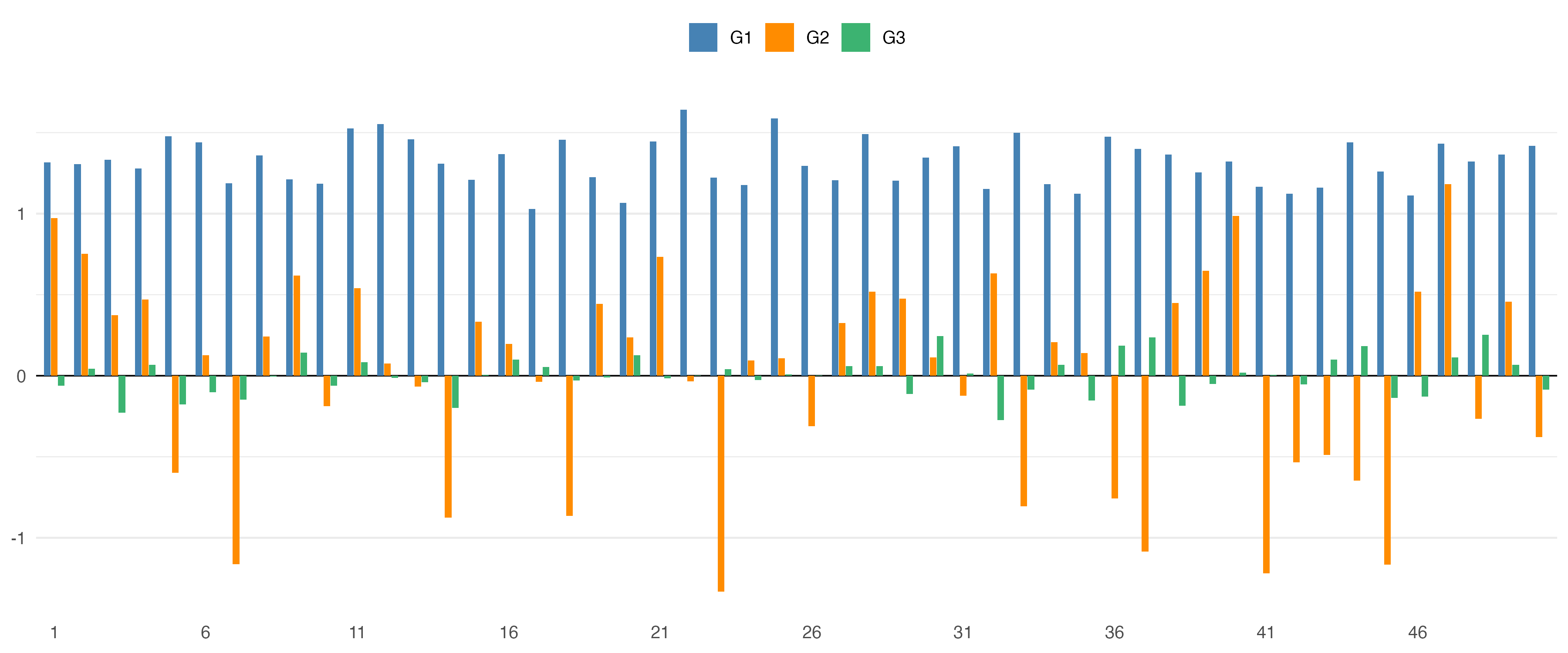}
\caption{Loadings of each of the first (blue), second (orange) and third (green) factors in the DFM with $N=50$.}
\label{fig:loadings_DFM}
\end{center}
\end{figure}

For comparison purposes, we also generate a standard DFM with $r = 3$ pervasive factors following the same design as in the ML-DFM experiments, without imposing zero restrictions on the loading matrix. The factors are extracted using the traditional PC estimator.\footnote{In this case, we align the estimated factors with the true ones applying the Procrustes rotation after factor extraction; see Gower (1975).} Figure~\ref{fig:loadings_DFM} reports the loadings of the three factors for $N = 50$. The first factor features large and positive loadings, while the second and third factors load with smaller magnitudes, with positive and negative signs. The loadings of the third factor are smaller than those of the second factor.

Using the factors, loadings, and idiosyncratic components described above, we generate $M=1000$ replications from each design considered. In each replication, the global and group-specific factors of the ML-DFM (the pervasive factors of the DFM) are estimated using the SLS (PC) procedure, obtaining $\hat{G}_t^{m}$, $\hat{L}_{1t}^{m}$, and $\hat{L}_{2t}^{m}$ ($\widetilde{F}_{1t}. \widetilde{F}_{2t}$, and $\widetilde{F}_{3t}$), for $t = 1,\ldots,T$. 
In addition, at each moment of time $t$, we calculate the following empirical MSE matrix based on the $M$ estimates of the factors
\begin{equation}
\label{eq:MC_MSE}
\mathbf{MSE}_t^{(emp)}=\frac{1}{M}\sum_{m=1}^M \left[ \left( \widetilde{\mathbf{F}}_t^{(m)} - \mathbf{F}_t \right) \left( \widetilde{\mathbf{F}}_t^{(m)} - \mathbf{F}_t \right)^{\prime} \right],
\end{equation}
where $\widetilde{\mathbf{F}}_t^{(m)}$ is the vector of estimates of the factors at time $t$, which can be $\widetilde{\mathbf{F}}^{(m)}_t= \left(\hat{G}_t^{m}, \hat{L}_{1t}^{m},\hat{L}_{2t}^{m} \right)^{\prime}$ when dealing with the ML-DFM or $\widetilde{\mathbf{F}}^{(m)}_t= \left( \widetilde{F}_{1t}. \widetilde{F}_{2t}, \widetilde{F}_{3t} \right)$ when dealing with the DFM. The empirical $3 \times 3$ MSE matrix in (\ref{eq:MC_MSE}) can be decomposed into the empirical covariance matrix and the product of the biases as follows:
\begin{equation}
\label{eq:MC_MSE_2}
\mathbf{MSE}_t^{(emp)}=\frac{1}{M}\sum_{m=1}^M \left[ \left(\widetilde{\mathbf{F}}_t^{(m)} - \overline{\widetilde{\mathbf{F}}}_t \right) \left(\widetilde{\mathbf{F}}_t^{(m)} - \overline{\widetilde{\mathbf{F}}}_t \right)^{\prime}+ \left( \mathbf{F}_t- \overline{\widetilde{\mathbf{F}}}_t \right) \left(\mathbf{F}_t^{(m)} - \overline{\widetilde{\mathbf{F}}}_t \right)^{\prime} \right],
\end{equation}
where the vector of Monte Carlo means of the factors is $\overline{\widetilde{\mathbf{F}}}_t=\frac{1}{M}\sum_{m=1}^M \widetilde{\mathbf{F}}_t^{(m)}$. 

The average (over time) of the empirical MSE matrix is, therefore, given by
\begin{equation}
\label{eq:emp}
\overline{\mathbf{MSE}}^{(emp)}=\frac{1}{T}\sum_{t=1}^T \mathbf{MSE}_t^{(emp)}.
\end{equation}

Second, for each design, we calculate the asymptotic MSE matrix as in (\ref{eq:avar}) using the true parameter values used to simulate $\mathbf{Y}_t$. By comparing the elements of this asymptotic MSE matrix with those of the empirical covariance matrix in (\ref{eq:emp}), we can assess whether the asymptotic distribution of the SLS (PC) factors extracted from the ML-DFM (DFM) matches that derived by Bai (2003) for PC factors extracted from a DFM, and therefore whether it delivers appropriate confidence regions in finite samples.

Third, for each design, replication, and time point, we estimate the asymptotic MSE matrix using the HR estimator based on (\ref{eq:gamma_1}) and the FPR estimator based on (\ref{eq:FPR}), both with and without subsampling correction. By comparing these estimated MSEs to the true ones computed using the true parameter values, we evaluate which estimator provides a more accurate approximation to the asymptotic MSE.

Consider first the results obtained for the DFM with $r = 3$ pervasive factors extracted using PC. Tables \ref{tab:MC_DFM3_1}--\ref{tab:MC_DFM3_3} report the results under three configurations for the idiosyncratic components: (i) cross-sectionally uncorrelated and homoscedastic idiosyncratic components; (ii) cross-sectionally uncorrelated and heteroscedastic idiosyncratic components; and (iii) heteroscedastic with weakly cross-correlated idiosyncratic components.

\begin{table}[ht!]
\centering
\begin{tabular}{|l|ccc|ccccc|}
\hline
& \multicolumn{3}{c}{Empirical} & \multicolumn{5}{c}{Asymptotic} \\
& MSE & Cov & Bias$^2$ & MSE  & HR & HRS & FPR & FPRS\\
\cline{2-9}
& \multicolumn{8}{c}{$N_=50,T=50$} \\
\cline{2-9}
$F_{1t}$ & 0.027 & 0.027 & 0.000 & 0.028  & 0.025& 0.026 & 0.024 & 0.025\\
$F_{2t}$ & 0.119 & 0.119 & 0.000 & 0.127  & 0.106 & 0.111 & 0.103 & 0.108\\
$F_{3t}$ & 3.730 & 3.378 & 0.351 & 3.483  & 1.674 & 1.858 & 1.616 & 1.800\\
$F_{1t}, F_{2t}$ & 0.000 & 0.000 & 0.000 & 0.000 & 0.001 & 0.000 & 0.001 & 0.000\\
$F_{2t}, F_{3t}$ & 0.002 & 0.002 & 0.000 & 0.000  & 0.001 & 0.002 & 0.001 & 0.002 \\
$F_{1t}, F_{3t}$ & 0.002 & 0.002 & 0.000 & 0.000  & 0.001 & 0.000 & 0.001 & 0.000\\
\cline{2-9}
& \multicolumn{8}{c}{$N=100,T=50$} \\
\cline{2-9}
$F_{1t}$ & 0.014 & 0.014 & 0.000 & 0.015  & 0.013 & 0.014 & 0.013 & 0.014\\
$F_{2t}$ & 0.067 & 0.067 & 0.000 & 0.071  & 0.062 & 0.066 & 0.062 & 0.065\\
$F_{3t}$  & 1.439 & 1.386 & 0.053 & 1.307 & 0.832 & 0.908 & 0.819 & 0.895\\
$F_{1t}, F_{2t}$  & 0.000 & 0.000 & 0.000 & 0.000 & 0.000 & 0.000 & 0.000 & 0.000 \\
$F_{2t}, F_{3t}$ & 0.000 & 0.000 & 0.000 & 0.000  & 0.000 & 0.000 & 0.000 & 0.000  \\
$F_{1t}, F_{3t}$ & 0.000 & 0.001 & 0.000 & 0.000  & 0.000 & 0.001 & 0.000 & 0.001\\
\cline{2-9}
& \multicolumn{8}{c}{$N=100,T=100$} \\
\cline{2-9}
$F_{1t}$  & 0.014 & 0.014 & 0.000 & 0.015 & 0.014 & 0.014 & 0.014 & 0.014\\
$F_{2t}$ & 0.068 & 0.068 & 0.000 & 0.071  & 0.065 & 0.069 & 0.065 & 0.069\\
$F_{3t}$ & 1.324 & 1.279 & 0.045  & 1.307 & 0.956 & 1.024 & 0.954 & 1.022\\
$F_{1t}, F_{2t}$ & 0.000 & 0.000 & 0.000 & 0.000  & 0.000 & 0.000 & 0.000 & 0.000 \\
$F_{2t}, F_{3t}$ & 0.001 & 0.001 & 0.000 & 0.000  & 0.000 & 0.000 & 0.000 & 0.000  \\
$F_{1t}, F_{3t}$  & -0.001 & -0.001 & 0.000 & 0.000 & 0.000 & 0.001 & 0.000 & 0.001\\
\cline{2-9}
& \multicolumn{8}{c}{$N=600,T=500$} \\
\cline{2-9}
$F_{1t}$  & 0.002 & 0.002 & 0.000 & 0.002 & 0.002 & 0.003 & 0.002 & 0.003\\
$F_{2t}$ & 0.012 & 0.012 & 0.000 & 0.013  & 0.012 & 0.013 & 0.012 & 0.013\\
$F_{3t}$  & 0.216 & 0.214 & 0.001 & 0.214 & 0.202 & 0.212 & 0.202 & 0.212\\
$F_{1t}, F_{2t}$  & 0.000 & 0.000 & 0.000 & 0.000 & 0.000 & 0.000 & 0.000 & 0.000 \\
$F_{2t}, F_{3t}$ & 0.000 & 0.000 & 0.000 & 0.000  & 0.000 & 0.000 & 0.000 & 0.000  \\
$F_{1t}, F_{3t}$ & 0.000 & 0.000 & 0.000 & 0.000  & 0.000 & 0.000 & 0.000 & 0.000\\
\hline
\end{tabular}
\caption{The table reports the elements of the average empirical MSE matrix, $(\overline{\mathbf{MSE}}^{(emp)})$, covariances, and cross-product bias matrices (first three columns), together with the finite-sample approximations of the elements of the asymptotic MSE matrix computed using the true model parameters and estimated using HR without and with subsampling correction (HRS), and FPR without and with subsampling correction (FPRS). The DGP corresponds to a DFM with cross-sectionally homoscedastic and uncorrelated idiosyncratic components. The factors are extracted using PC. All entries are multiplied by 10.} 
\label{tab:MC_DFM3_1}
\end{table}

Table \ref{tab:MC_DFM3_1} shows that, regardless of the cross-sectional and temporal dimensions, $N$ and $T$, when the idiosyncratic components are cross-sectionally homoscedastic and uncorrelated, and provided that the loadings are sufficiently large relative to the idiosyncratic variance, the empirical biases in the PC factor estimates are very small. However, when the loadings of a particular factor (the third) are not large enough, the estimates exhibit non-negligible biases that decrease with the cross-sectional dimension, $N$, and only slightly with the temporal dimension, $T$. In addition, the empirical MSE of each estimated PC factor decreases with $N$ and only marginally with $T$. This result confirms that the distribution of the estimated factors depends mainly on the cross-sectional dimension and marginally on the temporal dimension through the estimation of the loadings. Moreover, Table \ref{tab:MC_DFM3_1} shows that, regardless of the cross-sectional and temporal dimensions, the pairwise empirical correlations between the estimated factors are essentially zero.

Second, Table \ref{tab:MC_DFM3_1} shows that the elements of the asymptotic MSE matrix provide very accurate approximations to their empirical counterparts even when $N$ and $T$ are relatively small.

Third, Table \ref{tab:MC_DFM3_1} further shows that, for small samples, the HR estimator of the asymptotic MSE slightly underestimates the variances. Introducing the subsampling correction helps mitigate these negative biases in the estimation of the dispersion of PC factors. It should also be noted that the asymptotic MSE matrices estimated using HR and FPR are remarkably similar. When there is no cross-sectional correlation, there is no cost associated with using FPR instead of HR.

\begin{figure}[h!]
\begin{center}
\begin{subfigure}[b]{1\linewidth}
\includegraphics[trim=15mm 15mm 0mm 18mm, clip, width=1\textwidth]{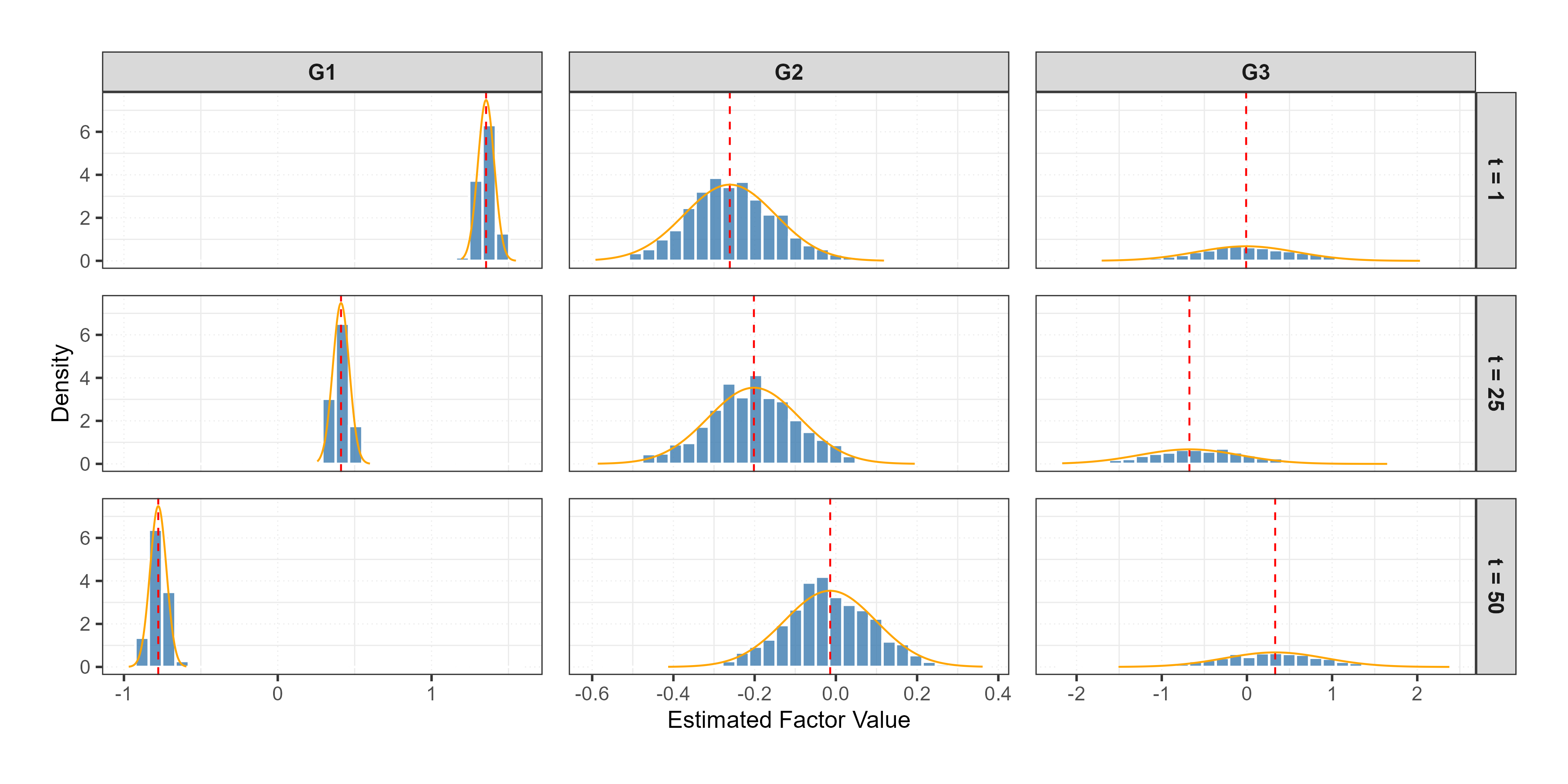}
\caption{$N_1=N_2=25, T=50$}
\end{subfigure}
\begin{subfigure}[b]{1\linewidth}
\includegraphics[trim=15mm 15mm 0mm 18mm, clip, width=1\textwidth]{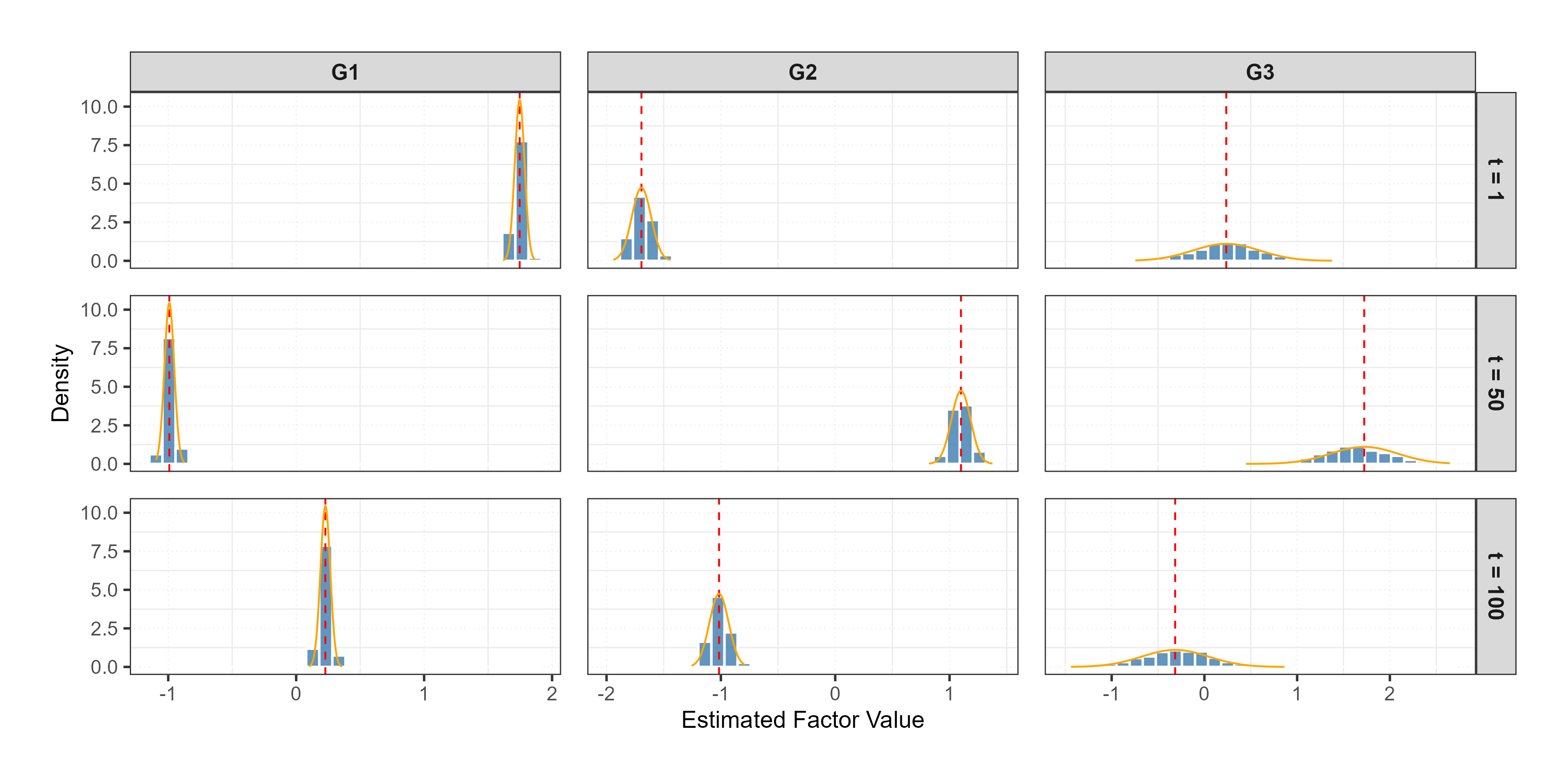}
\caption{$N_1=25, N_2=75, T=100$}
\end{subfigure}
\begin{subfigure}[b]{1\linewidth}
\includegraphics[trim=15mm 15mm 0mm 18mm, clip, width=1\textwidth]{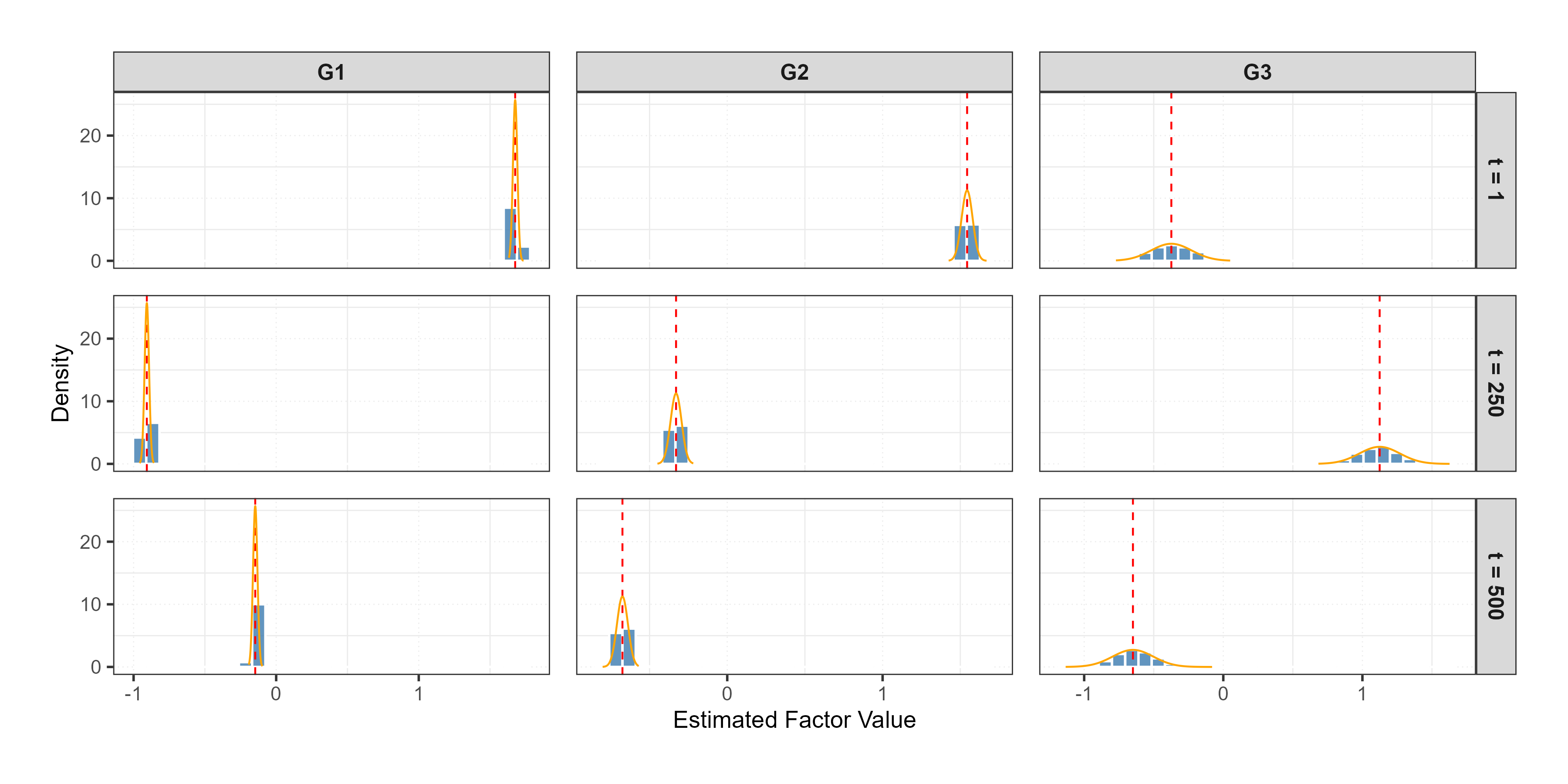}
\caption{$N_1=N_2=300, T=500$}
\end{subfigure}
\caption{Histograms of the estimated factors in a DFM with $r = 3$ at three time points: $t = 1$ (first column), $t = T/2$ (second column), and $t = T$ (third column), for the first factor (top row), second factor (middle row), and third factor (bottom row). The red vertical line indicates the true factor value. The orange curve represents the asymptotic density, computed using the MSE with the true parameter values. The idiosyncratic errors are cross-sectionally homoscedastic and uncorrelated. Each panel corresponds to a different $(N,T)$ configuration.}
\label{fig:densities_DFM3_1}
\end{center}
\end{figure}

Figure \ref{fig:densities_DFM3_1}, which plots histograms of each of the PC factors estimated through the Monte Carlo replicates at three particular moments of time,  together with their corresponding asymptotic densities computed using the true parameter values, shows that the latter provide an accurate approximation even when $N$ and $T$ are not very large.

Table \ref{tab:MC_DFM3_2} reports the results for the PC factors extracted from the DFM when the idiosyncratic components are cross-sectionally uncorrelated but heteroscedastic. We can observe that idiosyncratic heteroscedasticity has several relevant implications for the empirical distribution of the PC estimators. First, for small samples, both the empirical MSEs and the empirical biases are noticeably larger relative to the homoscedastic case, although both decrease with $N$ and $T$. This results are in concordance with those of Boivin and Ng (2006), who show that, for fixed $N$, the further the idiosyncratic components are from being homoscedastic and mutually uncorrelated, the less precise the PC estimator becomes. Bai and Li (2016) also argue that under fixed $N$, ignoring cross-sectional heteroscedasticity leads to inconsistent estimates of loadings due to incidental parameter bias; under large $N$, heteroscedasticity does not affect consistency, but still affects bias and efficiency; see also Stock and Watson (2002), who document deterioration in the performance of OLS-PC when the idiosyncratic components are strongly serially and/or cross-sectionally correlated, and to a lesser extent under heteroscedasticity. 
The empirical correlations between the estimated factors are also larger. For instance, in the design $N = T = 50$, the implied correlation between the first and second estimated factors is 0.087.\footnote{Note that the correlations can be easily computed from the reported covariances and MSEs.} These correlations decrease as the sample size increases. 

\begin{table}[ht!]
\centering
\begin{tabular}{l|ccc|ccccc}
\hline
& \multicolumn{3}{c}{Empirical} & \multicolumn{5}{c}{Asymptotic} \\
& MSE & Cov &Bias$^2$ & MSE & HR & HRS & FPR & FPRS\\
\cline{2-9}
& \multicolumn{8}{c}{$N=50,T=50$} \\
\cline{2-9}
$F_{1t}$ & 0.037 & 0.037 & 0.000 & 0.039  & 0.034 & 0.036 & 0.033 & 0.035\\
$F_{2t}$ & 0.172 & 0.172 & 0.001 & 0.181   & 0.149 & 0.157 & 0.144 & 0.152\\
$F_{3t}$ & 6.785 & 5.627 & 1.157 & 4.693  & 2.044 & 2.534 & 1.976 & 2.466\\
$F_{1t}, F_{2t}$ & 0.007 & 0.007 & 0.000 & 0.007  & 0.006 & 0.006 & 0.006 & 0.006\\
$F_{2t}, F_{3t}$  & 0.021 & 0.017 & 0.004 & 0.027 & 0.008 & 0.010 & 0.008 & 0.010 \\
$F_{1t}, F_{3t}$ & -0.002 & -0.001 & 0.000 & 0.018  & -0.001 & -0.002 & -0.001 & -0.002\\
\cline{2-9}
& \multicolumn{8}{c}{$N=100,T=50$} \\
\cline{2-9}
$F_{1t}$ & 0.018 & 0.018 & 0.000 & 0.019  & 0.017 & 0.018 & 0.017 & 0.018\\
$F_{2t}$ & 0.088 & 0.087 & 0.000 & 0.092  & 0.081 & 0.086 & 0.080 & 0.085\\
$F_{3t}$ & 2.037 & 1.931 & 0.106 & 1.608  & 1.016 & 1.133 & 1.002 & 1.118\\
$F_{1t}, F_{2t}$ & 0.001 & 0.001 & 0.000 & 0.001  & 0.001 & 0.001 & 0.001 & 0.000 \\
$F_{2t}, F_{3t}$ & -0.005 & -0.004 & 0.000 & -0.005 & -0.003 & -0.003 & -0.003 & -0.003  \\
$F_{1t}, F_{3t}$ & 0.011 & 0.010 & 0.001 & 0.013  & 0.006 & 0.008 & 0.006 & 0.008\\
\cline{2-9}
& \multicolumn{8}{c}{$N=100,T=100$} \\
\cline{2-9}
$F_{1t}$ & 0.019 & 0.019 & 0.000 & 0.019  & 0.018 & 0.019 & 0.018 & 0.019\\
$F_{2t}$ & 0.090 & 0.090 & 0.000 & 0.092  & 0.084 & 0.089 & 0.084 & 0.089\\
$F_{3t}$ & 1.767 & 1.688 & 0.080 & 1.608  & 1.174 & 1.267 & 1.172 & 1.265\\
$F_{1t}, F_{2t}$  & 0.001 & 0.001 & 0.000 & 0.001 & 0.000 & 0.000 & 0.000 & 0.000 \\
$F_{2t}, F_{3t}$  & -0.003 & -0.003 & 0.000 & -0.005 & -0.003 & -0.003 & -0.003 & -0.003  \\
$F_{1t}, F_{3t}$ & 0.011 & 0.010 & 0.001 & 0.013  & 0.008 & 0.010 & 0.008 & 0.010\\
\cline{2-9}
& \multicolumn{8}{c}{$N=600,T=500$} \\
\cline{2-9}
$F_{1t}$ & 0.003 & 0.003 & 0.000 & 0.003  & 0.003 & 0.003 & 0.003 & 0.003\\
$F_{2t}$  & 0.016 & 0.016 & 0.000 & 0.016 & 0.015 & 0.016 & 0.015 & 0.016\\
$F_{3t}$ & 0.261 & 0.259 & 0.002 & 0.256  & 0.242 & 0.255 & 0.242 & 0.255\\
$F_{1t}, F_{2t}$ & 0.000 & 0.000 & 0.000 & 0.000  & 0.000 & 0.000 & 0.000 & 0.000 \\
$F_{2t}, F_{3t}$ & 0.000 & 0.000 & 0.000 & 0.000  & 0.000 & 0.000 & 0.000 & 0.000  \\
$F_{1t}, F_{3t}$  & 0.000 & 0.000 & 0.000 & 0.000 & 0.000 & 0.000 & 0.000 & 0.000\\
\hline
\end{tabular}
\caption{The table reports the averages of empirical MSE $(\overline{\mathbf{MSE}}^{(emp)})$, covariances, and cross-product bias matrices (first three columns), together with the finite-sample approximations of the elements of the asymptotic MSE matrix computed using the true model parameters and estimated using HR without and with subsampling correction (HRS), and FPR without and with subsampling correction (FPRS). The DGP corresponds to a DFM with cross-sectionally heteroscedastic and uncorrelated idiosyncratic components. The factors are extracted using PC. All entries are multiplied by 10.} 
\label{tab:MC_DFM3_2}
\end{table}

Second, we can observe that because of the biases in the estimated factors, the asymptotic MSE matrices approximate the empirical covariance matrices rather than the empirical MSE matrices, which are substantially larger.\footnote{The histograms and asymptotic densities are similar to those in Figure \ref{fig:densities_DFM3_1} and are omitted for brevity.} Finally, the negative biases of the HR and FPR estimators of the asymptotic MSE are more pronounced than in the homoscedastic case, making the subsampling correction particularly relevant in this setting.


Table \ref{tab:MC_DFM3_3}, which reports the results when the idiosyncratic components are also cross-sectionally correlated, shows that the empirical MSEs and biases become even larger. The conclusions regarding the ability of the asymptotic MSE to approximate the empirical distribution, as well as the conclusions about the performance of the asymptotic MSE estimators, remain broadly similar to those obtained under cross-sectional uncorrelatedness. The only additional conclusion is that, in this setting, the FPR estimator with subsampling delivers the most accurate estimates of the asymptotic MSE matrix.

\begin{table}[ht!]
\centering
\begin{tabular}{l|ccc|ccccc}
\hline
& \multicolumn{3}{c}{Empirical} & \multicolumn{5}{c}{Asymptotic} \\
& MSE & Cov & Bias$^2$ & MSE & HR & HRS & FPR & FPRS\\
\cline{2-9}
& \multicolumn{8}{c}{$N=50,T=50$} \\
\cline{2-9}
\\
$F_{1t}$ & 0.014 & 0.014 & 0.000 & 0.015  & 0.034 & 0.036 & 0.028 & 0.030\\
$F_{2t}$  & 0.186 & 0.184 & 0.001 & 0.193 & 0.154 & 0.163 & 0.145 & 0.155\\
$F_{3t}$ & 12.128 & 8.435 & 3.693 & 5.236  & 1.569 & 2.273 & 1.598 & 2.301\\
$F_{1t}, F_{2t}$ & -0.003 & -0.003 & 0.000 & -0.003  & 0.002 & 0.001 & 0.001 & 0.000\\
$F_{2t}, F_{3t}$  & 0.002 & 0.002 & 0.001& -0.006  & -0.003 & -0.003 & -0.002 & -0.002 \\
$F_{1t}, F_{3t}$ & 0.062 & 0.042 & 0.019 & 0.208  & -0.009 & -0.012 & -0.006 & -0.008\\
\cline{2-9}
& \multicolumn{8}{c}{$N=100,T=50$} \\
\cline{2-9}
\\
$F_{1t}$& 0.007 & 0.007 & 0.000  & 0.007  & 0.017 & 0.018 & 0.011 & 0.012\\
$F_{2t}$ & 0.101 & 0.100 & 0.000 & 0.104 & 0.081 & 0.085 & 0.080 & 0.085\\
$F_{3t}$ & 3-519 & 3.205 & 0.313 & 1.784  & 0.972 & 1.188 & 1.060 & 1.276\\
$F_{1t}, F_{2t}$ & -0.001 & -0.001 & 0.000 & -0.001  & -0.001 & -0.001 & -0.001 & -0.001 \\
$F_{2t}, F_{3t}$ & 0.002 & 0.002 & 0.000 & 0.002  & 0.000 & 0.000 & 0.000 & 0.000  \\
$F_{1t}, F_{3t}$ & -0.026 & -0.023 & -0.002 & -0.024  & 0.000 & 0.001 & -0.004 & -0.003\\
\cline{2-9}
& \multicolumn{8}{c}{$N_1=100,T=100$} \\
\cline{2-9}
\\
$F_{1t}$ & 0.007 & 0.007 & 0.000  & 0.007 & 0.018 & 0.019 & 0.006 & 0.007\\
$F_{2t}$ & 0.103 & 0.102 & 0.000 & 0.104  & 0.084 & 0.089 & 0.087 & 0.091\\
$F_{3t}$ & 2.761 & 2.567 & 0.194 & 1.784  & 1.130 & 1.264 & 1.344 & 1.479\\
$F_{1t}, F_{2t}$  & -0.001& -0.001 & 0.000 & -0.001 & -0.001 &-0.001 & 0.000 & -0.001\\
$F_{2t}, F_{3t}$  & 0.003 & 0.002 & 0.000 & 0.002 & 0.000 & 0.000 & -0.001 & 0.000 \\
$F_{1t}, F_{3t}$  & -0.029 & -0.027 & -0.002 & -0.024 & -0.001 & 0.000 & -0.012 & -0.011  \\
\cline{2-9}
& \multicolumn{8}{c}{$N=600,T=500$} \\
\cline{2-9}
\\
$F_{1t}$ & 0.001 & 0.001 & 0.000 & 0.001  & 0.003 & 0.003 & 0.001 & 0.002\\
$F_{2t}$  & 0.015 & 0.015 & 0.000 & 0.015& 0.015 & 0.016 & 0.015 & 0.016\\
$F_{3t}$ & 0.277 & 0.275 & 0.002 & 0.258  & 0.250 & 0.264 & 0.257 & 0.270\\
$F_{1t}, F_{2t}$ & 0.000 & 0.000 & 0.000 & 0.000 & 0.000 & 0.000 & 0.000 & 0.000 \\
$F_{2t}, F_{3t}$  & -0.001 & -0.001 & 0.000 & 0.000 & 0.000 & 0.000 & -0.001 & 0.000  \\
$F_{1t}, F_{3t}$  & 0.003 & 0.003 & 0.000 & -0.001 & -0.001 & -0.001 & 0.003 & 0.003\\
\hline
\end{tabular}
\caption{The table reports the averages of empirical MSE $(\overline{\mathbf{MSE}}^{(emp)})$, covariances, and cross-product bias matrices (first three columns), together with the finite sample approximation of the elements of the Asymptotic MSE matrix obtained with the true models parameters and estimated using HR without and with subsampling correction (HRS), and FPR without and with subsampling correction (FPRS). The DGP is a DFM with cross-sectionally heteroscedastic and correlated idiosyncratic components. The factors are extracted using PC. All entries are multiplied by 10.} 
\label{tab:MC_DFM3_3}
\end{table}

\begin{figure}[h!]
\begin{center}
\begin{subfigure}[b]{1\linewidth}
\includegraphics[trim=15mm 15mm 0mm 18mm, clip, width=1\textwidth]{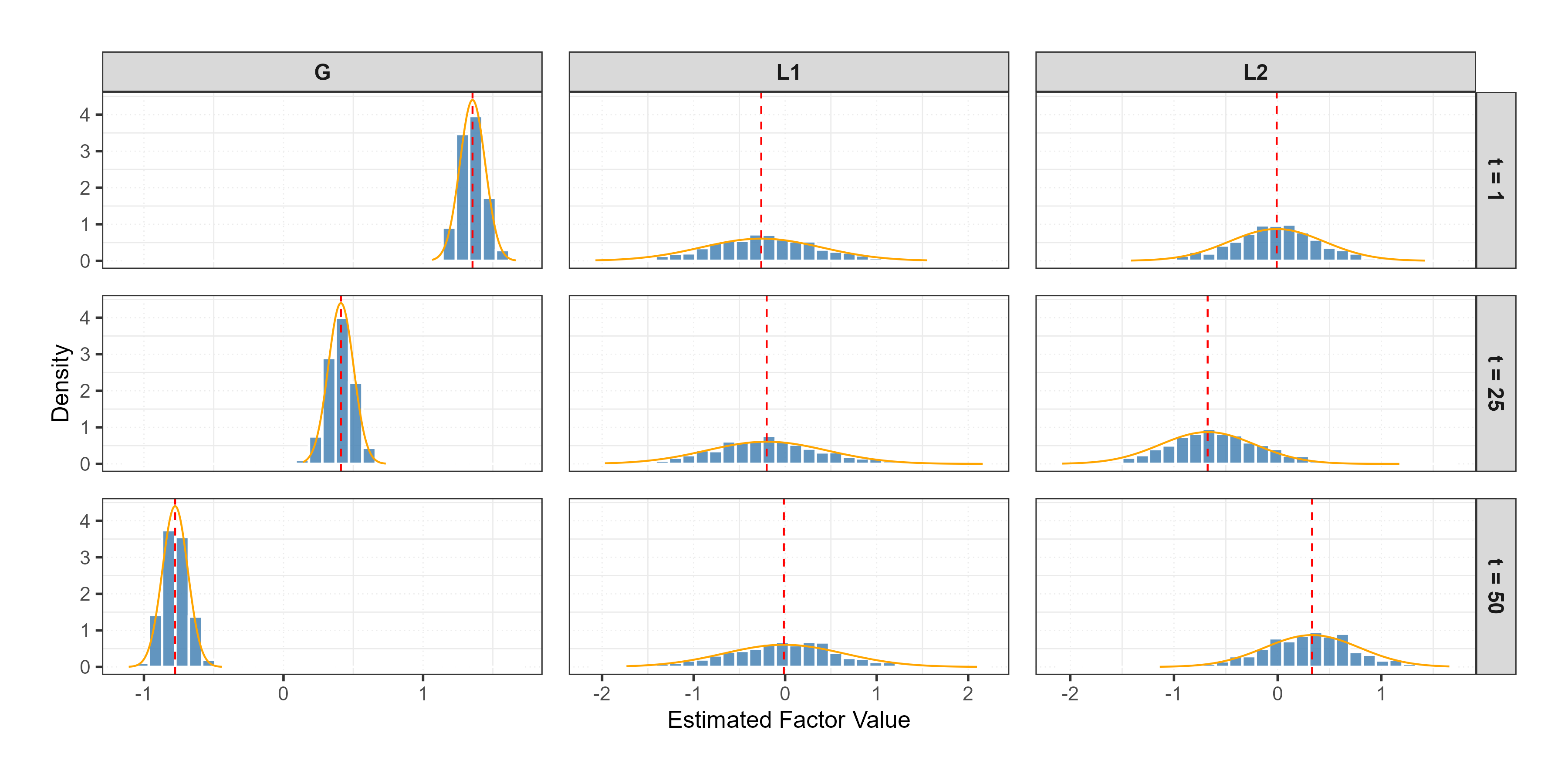}
\caption{$N_1=N_2=25, T=50$}
\end{subfigure}
\begin{subfigure}[b]{1\linewidth}
\includegraphics[trim=15mm 15mm 0mm 18mm, clip, width=1\textwidth]{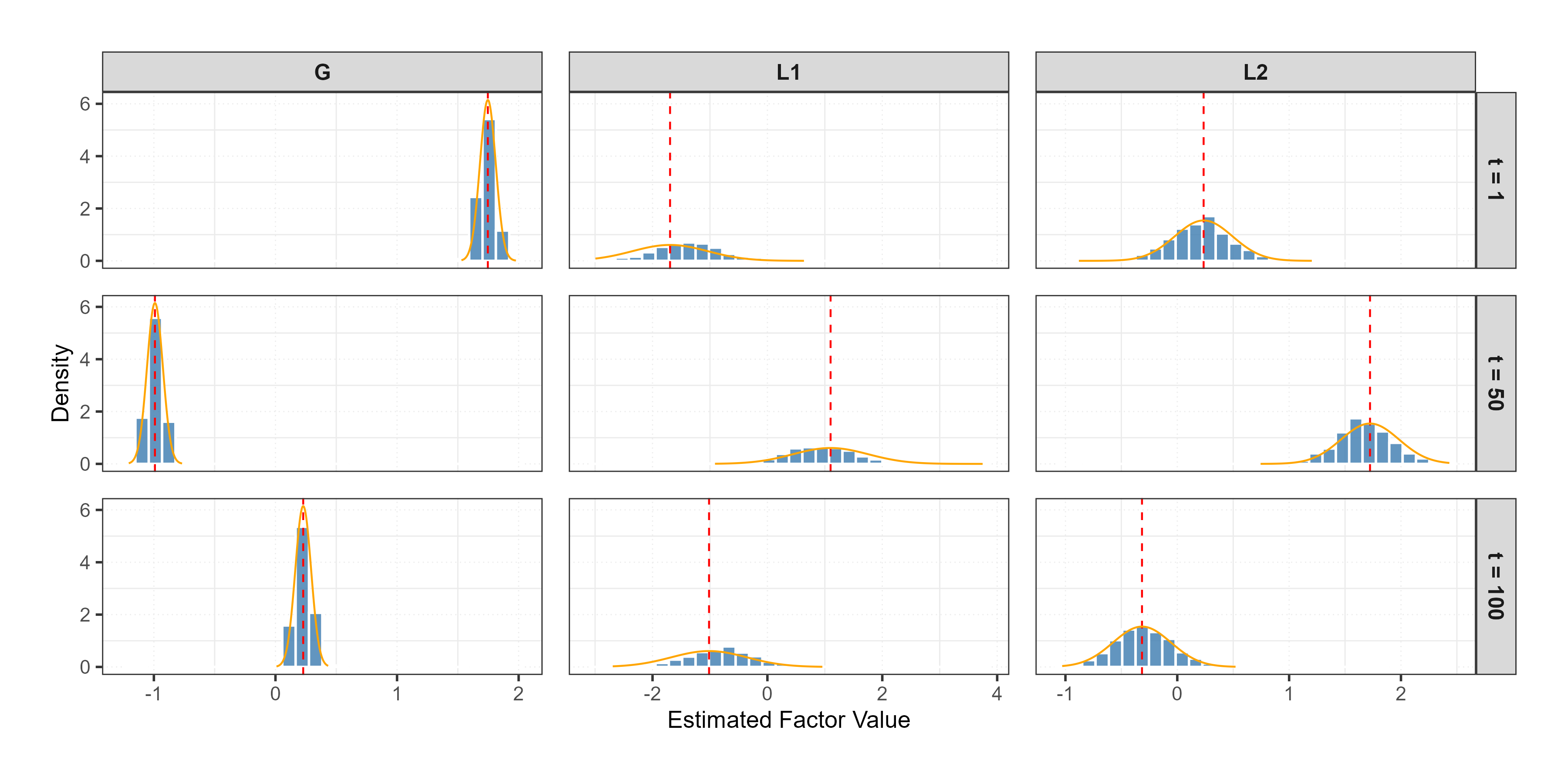}
\caption{$N_1=25, N_2=75, T=100$}
\end{subfigure}
\begin{subfigure}[b]{1\linewidth}
\includegraphics[trim=15mm 15mm 0mm 18mm, clip, width=1\textwidth]{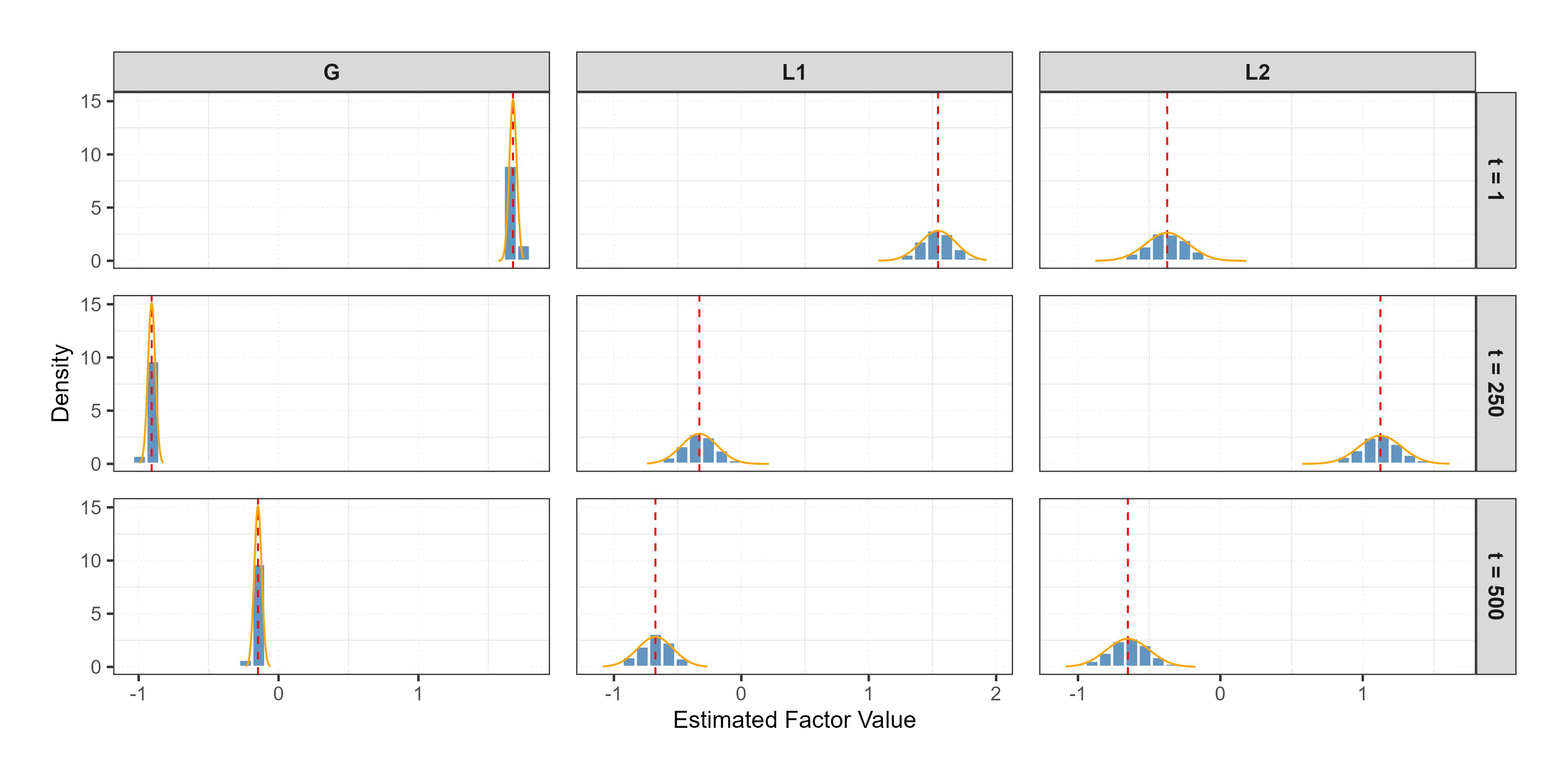}
\caption{$N_1=N_2=300, T=500$}
\end{subfigure}
\caption{Histograms of the estimated global factor, $G_t$ (first column), specific factor of the first block $L_{1t}$ (second column), and specific factor of the second block, $L_{2t}$ (third column), at three particular moments of time: i) $t=1$ (first column); ii) $t=T/2$ (second column); and iii) $t=T$ (third column). The red vertical line represents the true factor's value. The idiosyncratic errors are cross-sectionally homoscedastic and uncorrelated.}
\label{fig:densities_1}
\end{center}
\end{figure}

Consider now the results for the SLS estimator of the factors (one global factor $G_t$ and one group-specific factor per block, $L_{1t}$ and $L_{2t}$) in the ML-DFM. Table \ref{tab:MC_1} reports the Monte Carlo averages of the elements of the empirical MSE matrix together with their decomposition into the corresponding elements of the empirical covariance matrix and the cross-product of the empirical biases, when the idiosyncratic components are cross-sectionally homoscedastic and uncorrelated. Comparing these results to those reported in Table \ref{tab:MC_DFM3_1} for the PC factors extracted from the DFM, we observe that the two sets of results are very similar. The empirical biases are small unless the sample size is small and the loadings are weak. Regarding the estimation of the asymptotic MSE, we obtain essentially the same results regardless of whether the HR or FPR estimators of the covariance matrix are used, as expected given that the idiosyncratic components are cross-sectionally uncorrelated. Moreover, the subsampling correction is particularly relevant for accounting for uncertainty in the estimation of loadings when $T$ is not sufficiently large. Even more important for the main goal of this paper is to observe that, for moderate sample sizes, the asymptotic MSEs provide accurate approximations to the empirical MSEs. Similarly, Figure \ref{fig:densities_1}, which plots the empirical finite-sample densities together with the corresponding asymptotic normal densities, shows that the asymptotic distribution derived by Bai (2003) for PC factors in DFMs provides a very accurate approximation of the finite sample approximation of the SLS estimator of the factors in ML-DFMs. Consequently, the asymptotic distribution of the SLS estimator can be characterised as in (\ref{eq:asymptotic_1}), although with slower rates of convergence; see also Tu and Zheng (2025), who document similar behaviour.

\begin{table}[ht!]
\centering
\begin{tabular}{l|ccc|ccccc}
\hline
& \multicolumn{3}{c}{Empirical} & \multicolumn{5}{c}{Asymptotic} \\
 & MSE & Cov & Bias$^2$ & MSE & HR & HRS & FPR & FPRS\\
\cline{2-9}
& \multicolumn{8}{c}{$N_1=25, N_2=25,T=50$} \\
\cline{2-9}
$G_t$  & 0.087 & 0.086 & 0.000 & 0.082 & 0.074 & 0.077 & 0.071 & 0.075\\
$L_{1t}$ & 4.035 & 3.622 & 0.412 & 4.319  & 2.124 & 2.306 & 2.052 & 2.234\\
$L_{2t}$  & 1.983 & 1.883 & 0.100 & 2.099 & 1.360 & 1.439 & 1.315 & 1.393\\
$G_t, L_{1t}$ & -0.004 & -0.005 & 0.001  & 0.000 & -0.001 & -0.001 & -0.001 & -0.001 \\
$G_t, L_{2t}$  & 0.000 & -0.001 & 0.000 & 0.000 & 0.001 & 0.002 & -0.001 & 0.002  \\
$L_{1t}, L_{2t}$  & -0.006 & -0.006 & 0.000 & 0.000 & -0.001 & -0.001 & -0.001 & -0.001\\
\cline{2-9}
& \multicolumn{8}{c}{$N_1=25, N_2=75,T=50$} \\
\cline{2-9}
$G_t$  & 0.045 & 0.045 & 0.000 & 0.042 & 0.039 & 0.041 & 0.038 & 0.041\\
$L_{1t}$  & 3.981 & 3.581 & 0.400 & 4.319 & 2.144 & 2.317 & 2.103 & 2.276\\
$L_{2t}$  & 0.683 & 0.671 & 0.012 & 0.669 & 0.526 & 0.567 & 0.518 & 0.559\\
$G_t, L_{1t}$ & 0.003 & 0.003 & -0.001 & 0.000  & 0.001 & 0.001 & 0.001 & 0.001\\
$G_t, L_{2t}$  & 0.000 & 0.000 & 0.000 & 0.000 & 0.000 & 0.001 & 0.000 & 0.001\\
$L_{1t}, L_{2t}$  & -0.005 & -0.003 & -0.002 & 0.000  & 0.000 & 0.000 & 0.000 & 0.000\\
\cline{2-9}
& \multicolumn{8}{c}{$N_1=25, N_2=75,T=100$} \\
\cline{2-9}
$G_t$ & 0.043 & 0.043 & 0.000 & 0.042  & 0.040 & 0.042 & 0.040 & 0.042\\
$L_{1t}$ & 3.625 & 3.293 & 0.333 & 4.319  & 2.399 & 2.534 & 2.384 & 2.518\\
$L_{2t}$ & 0.660 & 0.649 & 0.011 & 0.669 & 0.562 & 0.602 & 0.562 & 0.602\\
$G_t, L_{1t}$  & 0.000 & 0.001 & -0.001 & 0.000 & 0.001 & 0.002 & 0.002 & 0.002 \\
$G_t, L_{2t}$ & 0.001 & 0.001 & 0.000 & 0.000  & 0.001 & 0.001 & 0.001 & 0.001  \\
$L_{1t}, L_{2t}$ & 0.006 & 0.006 & 0.000 & 0.000  & 0.000 & 0.000 & 0.000 & 0.000\\
\cline{2-9}
& \multicolumn{8}{c}{$N_1=300, N_2=300,T=500$} \\
\cline{2-9}
$G_t$  & 0.007 & 0.007 & 0.000 & 0.007 & 0.007 & 0.007 & 0.007 & 0.007\\
$L_{1t}$  & 0.198 & 0.197 & 0.001 & 0.199 & 0.191 & 0.201 & 0.191 & 0.201\\
$L_{2t}$  & 0.226 & 0.225 & 0.002 & 0.228 & 0.217 & 0.228 & 0.217 & 0.228\\
$G_t, L_{1t}$ & 0.000 & 0.000 & 0.000  & 0.000 & 0.000 & 0.000 & 0.000 & 0.000 \\
$G_t, L_{2t}$ & 0.000 & 0.000 & 0.000  & 0.000 & 0.000 & 0.000 & 0.000 & 0.000  \\
$L_{1t}, L_{2t}$ & 0.000 & 0.000 & 0.000 & 0.000   & 0.000 & 0.000 & 0.000 & 0.000\\
\hline
\end{tabular}
\caption{The table reports the averages of empirical MSE $(\overline{\mathbf{MSE}}^{(emp)})$, covariances, and cross-product biases matrices (first three columns), together with the finite-sample approximations of the elements of the asymptotic MSE matrix computed using the true model parameters and estimated using HR without and with subsampling correction (HRS), and FPR without and with subsampling correction (FPRS). The DGP corresponds to a ML-DFM with one global factor and one group-specific factor for each of the two groups of variables, and cross-sectionally homoscedastic and uncorrelated idiosyncratic components. The factors are extracted using SLS. All entries are multiplied by 10.
} 
\label{tab:MC_1}
\end{table}

To assess the effect of idiosyncratic cross-sectional heteroscedasticity on the previous conclusions, Table \ref{tab:MC_2} reports the results obtained when the idiosyncratic components are cross-sectionally uncorrelated but heteroscedastic. As in the case of the DFM, idiosyncratic heteroscedasticity increases the empirical MSE of the factors and deteriorates the performance of the asymptotic approximation of their MSEs, which requires large samples (in both $N$ and $T$) to be reliable. In addition to increasing uncertainty, heteroscedasticity induces sizable biases that decrease with both $N$ and $T$. These biases primarily affect the group-specific factors because, as shown in Figure \ref{fig:loadings_MLDFM}, their loadings are small in absolute value relative to those of the global factor.

\begin{figure}[h!]
\begin{center}
\begin{subfigure}[b]{1\linewidth}
\includegraphics[trim=15mm 15mm 0mm 18mm, clip, width=1\textwidth]{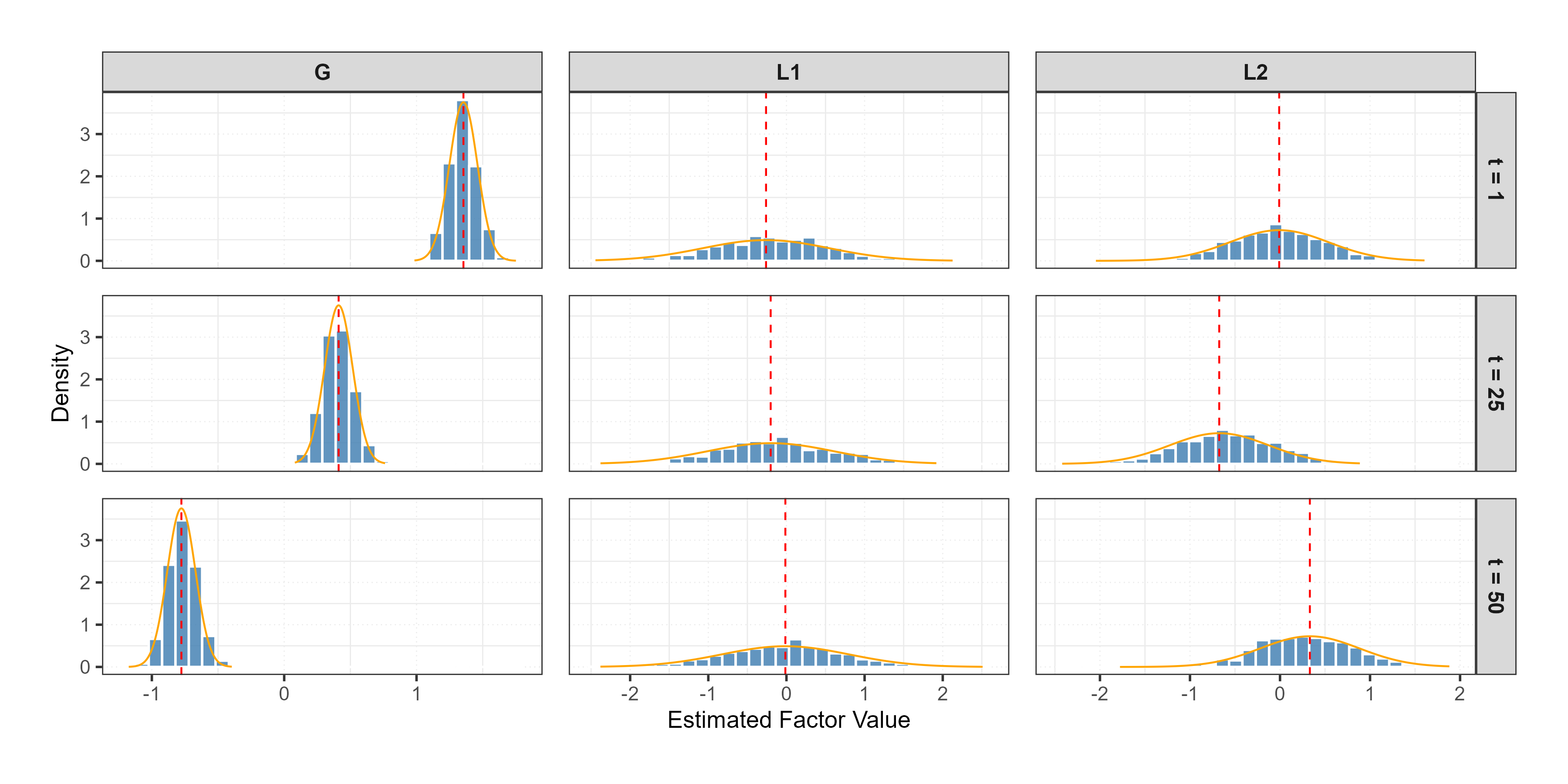}
\caption{$N_1=N_2=25, T=50$}
\end{subfigure}
\begin{subfigure}[b]{1\linewidth}
\includegraphics[trim=15mm 15mm 0mm 18mm, clip, width=1\textwidth]{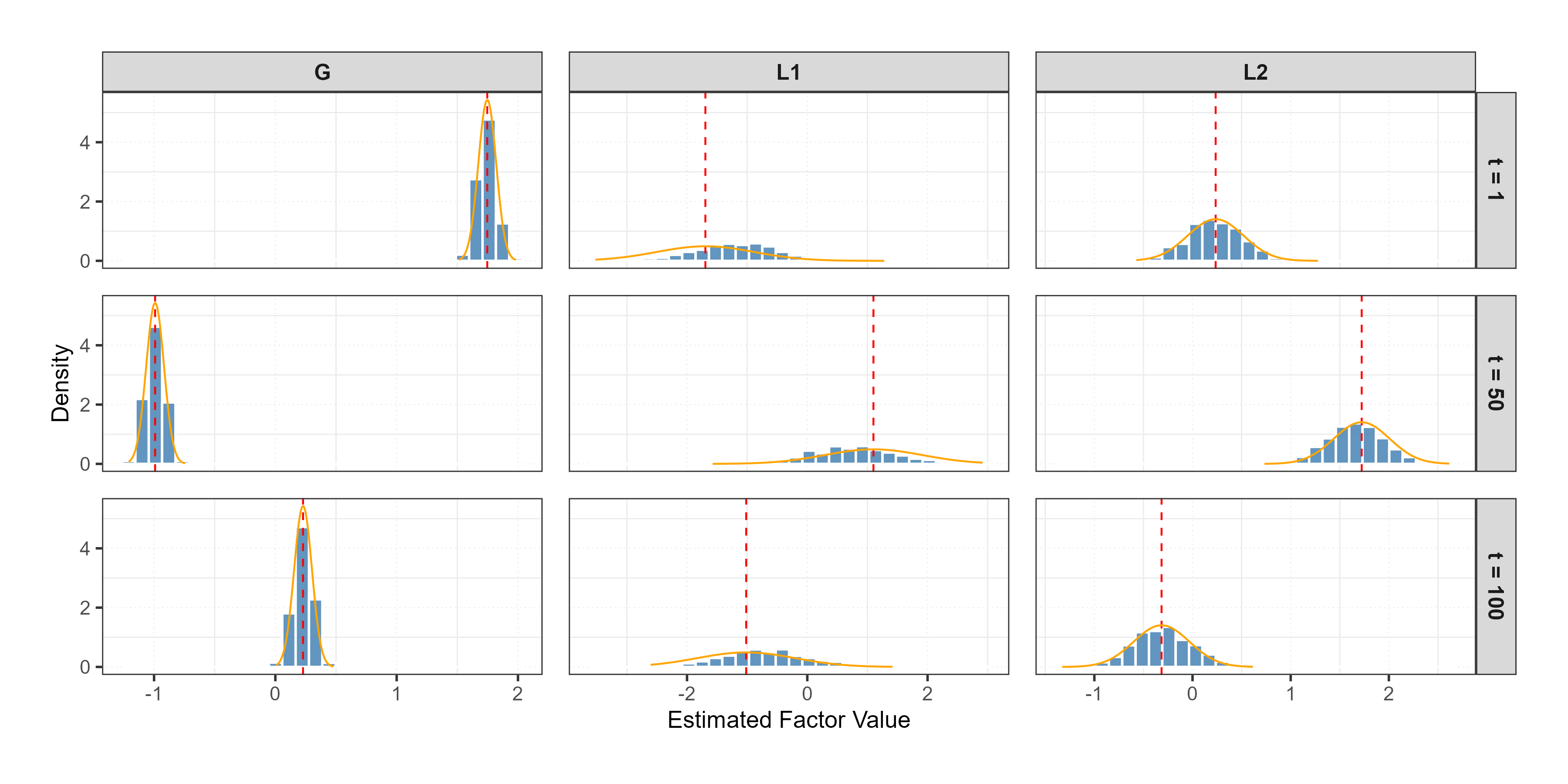}
\caption{$N_1=25, N_2=75, T=100$}
\end{subfigure}
\begin{subfigure}[b]{1\linewidth}
\includegraphics[trim=15mm 15mm 0mm 18mm, clip, width=1\textwidth]{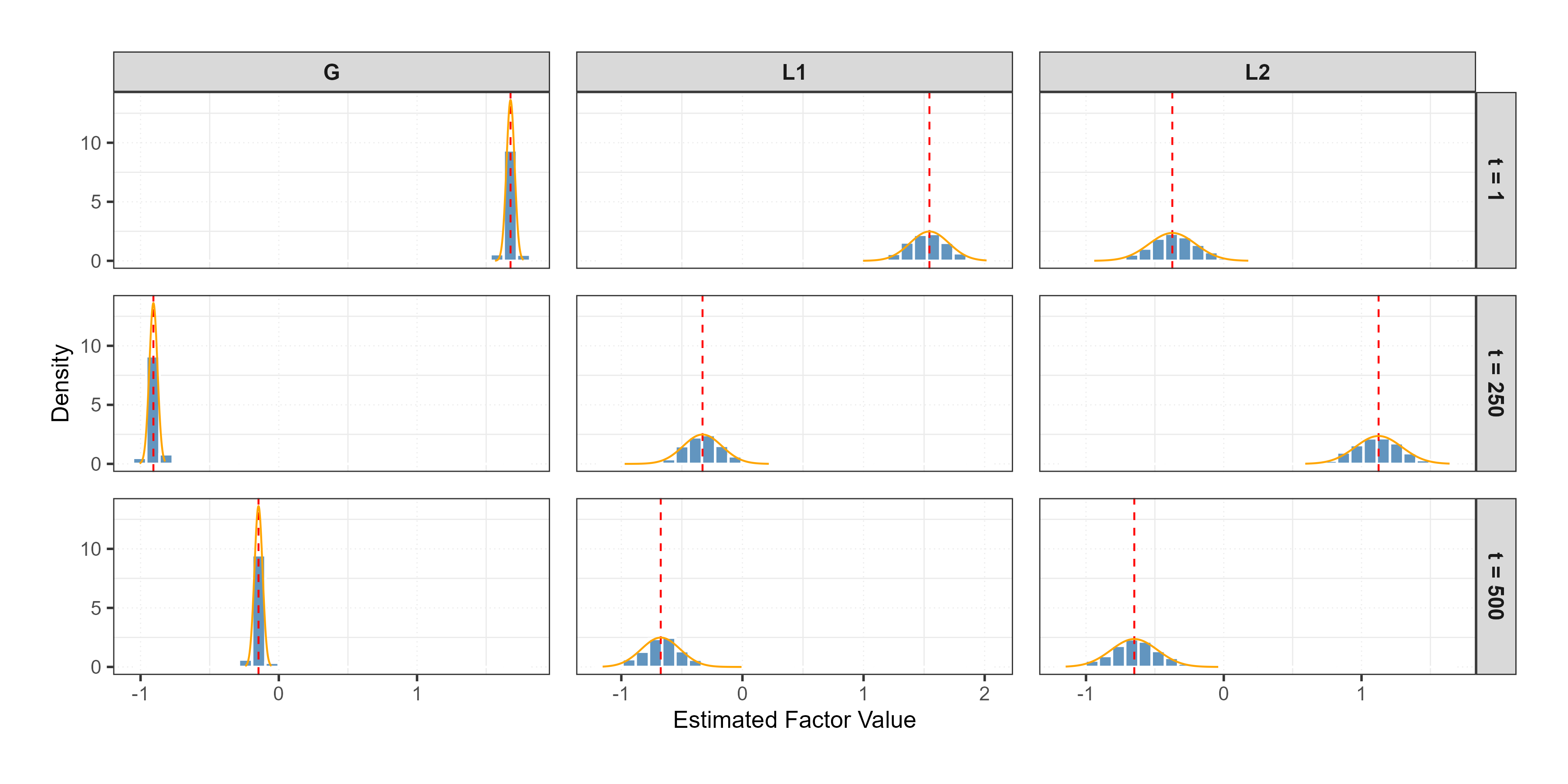}
\caption{$N_1=N_2=300, T=500$}
\end{subfigure}
\caption{Histograms of the estimated global factor, $G_t$ (first column), specific factor of the first block $L_{1t}$ (second column), and specific factor of the second block, $L_{2t}$ (third column), at three particular moments of time: i) $t=1$ (first column); ii) $t=T/2$ (second column); and iii) $t=T$ (third column). The red vertical line represents the true factor's value. The idiosyncratic errors are cross-sectionally uncorrelated and heteroscedastic.}
\label{fig:densities_2}
\end{center}
\end{figure}

Figure \ref{fig:densities_2}, which plots the histograms of the Monte Carlo SLS estimates of the factors at three time points together with the corresponding normal asymptotic densities, illustrates how the presence of biases can affect the estimation of the first group-specific factor when $N_1 = N_2 = 25$ and $T = 50$. Very large sample sizes are required for the bias to become negligible. Finally, as in the homoscedastic case, the MSE matrices estimated using HR and FPR are very similar, with subsampling playing an important role.

\begin{table}[ht!]
\centering
\begin{tabular}{l|ccc|ccccc}
\hline
& \multicolumn{3}{c}{Empirical} & \multicolumn{5}{c}{Asymptotic} \\
& MSE & Cov & Bias$^2$ & MSE & HR & HRS & FPR & FPRS\\
\cline{2-9}
& \multicolumn{8}{c}{$N_1=25, N_2=25,T=50$} \\
\cline{2-9}
$G_t$  & 0.121 & 0.121 & 0.001 & 0.113 & 0.101 & 0.107 & 0.097 & 0.103\\
$L_{1t}$  & 6.939 & 5.731 & 1.208 & 6.624 & 2.641 & 3.072 & 2.551 & 2.982\\
$L_{2t}$  & 2.961 & 2.739 & 0.222 & 3.023 & 1.805 & 1.929 & 1.735 & 1.895\\
$G_t, L_{1t}$  & 0.002 & 0.002 & -0.001 & 0.002 & 0.004 & 0.004 & 0.003 & 0.004\\
$G_t, L_{2t}$ & -0.016 & -0.013 & -0.003 & -0.026  & -0.006 & -0.006 & -0.006 & -0.005 \\
$L_{1t}, L_{2t}$  & 0.048 & 0.040 & 0.008 & 0.000 & 0.001 & 0.001 & 0.001 & 0.002\\
\cline{2-9}
& \multicolumn{8}{c}{$N_1=25, N_2=75,T=50$} \\
\cline{2-9}
$G_t$  & 0.058 & 0.058 & 0.000 & 0.054 & 0.050 & 0.053 & 0.049 & 0.052\\
$L_{1t}$  & 7.009 & 5.778 & 1.232 & 6.624& 2.662 & 3.069 & 2.586 & 2.993\\
$L_{2t}$ & 0.889 & 0.868 & 0.021& 0.807  & 0.639 & 0.694 & 0.630 & 0.684\\
$G_t, L_{1t}$ & -0.003 & -0.003 & 0.000& 0.071  & 0.000 & 0.000 & 0.000 & 0.000 \\
$G_t, L_{2t}$ & 0.006 & 0.005 & 0.001& 0.004  & 0.003 & 0.003 & 0.003 & 0.003  \\
$L_{1t}, L_{2t}$ & 0.004 & 0.008 & -0.003& 0.003  & 0.000 & 0.000 & 0.000 & 0.000\\
\cline{2-9}
& \multicolumn{8}{c}{$N_1=25, N_2=75,T=100$} \\
\cline{2-9}
$G_t$  & 0.056 & 0.056 & 0.000& 0.054 & 0.051 & 0.054 & 0.051 & 0.054\\
$L_{1t}$  & 5.622 & 4.824 & 0.797& 6.624 & 3.120 & 3.374 & 3.070 & 3.324\\
$L_{2t}$ & 0.828 & 0.810 & 0.018 & 0.807 & 0.684 & 0.735 & 0.684 & 0.735\\
$G_t, L_{1t}$ & 0.071 & 0.001 & 0.000& 0.001  & 0.002 & 0.002 & 0.001 & 0.002 \\
$G_t, L_{2t}$ & 0.004 & 0.003 & 0.000 & 0.004 & 0.004 & 0.004 & 0.004 & 0.004  \\
$L_{1t}, L_{2t}$  & -0.004 & -0.001 & -0.003& 0.000 & 0.000 & 0.000 & 0.000 & 0.000\\
\cline{2-9}
& \multicolumn{8}{c}{$N_1=300, N_2=300,T=500$} \\
\cline{2-9}
$G_t$ & 0.009 & 0.009 & 0.000 & 0.009 & 0.008 & 0.009 & 0.008 & 0.009\\
$L_{1t}$  & 0.257 & 0.255 & 0.002& 0.258 & 0.246 & 0.259 & 0.246 & 0.259\\
$L_{2t}$  & 0.285 & 0.283 & 0.002& 0.286 & 0.271 & 0.285 & 0.271 & 0.285\\
$G_t, L_{1t}$ & 0.000 & 0.000& 0.000 & 0.000 & 0.000 & 0.000 & 0.000 & 0.000 \\
$G_t, L_{2t}$ & 0.000 & 0.000 & 0.000 & 0.000 & 0.000 & 0.000 & 0.000 & 0.000  \\
$L_{1t}, L_{2t}$  & -0.001 & -0.001 & 0.000& 0.000 & 0.000 & 0.000 & 0.000 & 0.000\\
\hline
\end{tabular}
\caption{The table reports the averages of empirical MSE $(\overline{\mathbf{MSE}}^{(emp)})$, covariances, and cross-product biases matrices (first three columns), together with the finite-sample approximations of the elements of the asymptotic MSE matrix computed using the true model parameters and estimated using HR without and with subsampling correction (HRS), and FPR without and with subsampling correction (FPRS). The DGP corresponds to an ML-DFM with one global factor and one group-specific factor for each of the two groups of variables, and cross-sectionally heteroscedastic and uncorrelated idiosyncratic components. The factors are extracted using SLS. All entries are multiplied by 10.} 
\label{tab:MC_2}
\end{table}

Finally, Table \ref{tab:MC_3} reports the results when the idiosyncratic components of the ML-DFM are cross-sectionally heteroscedastic and correlated. We observe substantial biases in the estimation of the group-specific factors, consistent with the fact that their loadings are small in absolute value. Due to these biases, the asymptotic MSE matrix provides a better approximation to the finite-sample empirical covariance matrix than to the empirical MSE matrix. In line with the results for the DFM, the best estimator of the asymptotic MSE matrix in this setting is FPR, which accounts for cross-sectional correlation in the idiosyncratic components and benefits from subsampling to accommodate the uncertainty associated with the estimation of the loadings.

\begin{table}[ht!]
\centering
\begin{tabular}{l|ccc|ccccc}
\hline
& \multicolumn{3}{c}{Empirical} & \multicolumn{5}{c}{Asymptotic} \\
 & MSE & Cov & Bias$^2$& MSE & HR & HRS & FPR & FPRS\\
\cline{2-9}
& \multicolumn{8}{c}{$N_1=25, N_2=25,T=50$} \\
\cline{2-9}
$G_t$  & 0.065 & 0.065 & 0.000& 0.046 & 0.105 & 0.112 & 0.089 & 0.096\\
$L_{1t}$ & 10.391 & 7.678 & 2.713 & 6.604 & 2.110 & 2.693 & 2.046 & 2.629\\
$L_{2t}$  & 3.692 & 3.346 & 0.346& 3.055 & 1.753 & 1.936 & 1.690 & 1.873\\
$G_t, L_{1t}$  & 0.023 & 0.037 & -0.014 & 0.032 & 0.012 & 0.014 & 0.010 & 0.013\\
$G_t, L_{2t}$ & -0.017 & -0.014 & -0.003& -0.022  & 0.009 & 0.013 & 0.004 & 0.007 \\
$L_{1t}, L_{2t}$  & -0.307 & -0.206 & -0.100 & -0.642 & -0.002 & 0.012 & -0.016 & -0.002\\
\cline{2-9}
& \multicolumn{8}{c}{$N_1=25, N_2=75,T=50$} \\
\cline{2-9}
$G_t$  & 0.025 & 0.025 & 0.000& 0.020 & 0.049 & 0.053 & 0.034 & 0.037\\
$L_{1t}$ & 6.874 & 5.685 & 1.188& 6.528  & 2.237 & 2.598 & 2.159 & 2.520\\
$L_{2t}$ & 1.220 & 1.181 & 0.038& 0.983  & 0.665 & 0.729 & 0.688 & 0.752\\
$G_t, L_{1t}$ & 0.000 & -0.003 & 0.003 & 0.008 & 0.006 & 0.006 & 0.002 & 0.002 \\
$G_t, L_{2t}$ & -0.026 & -0.028 & 0.002& -0.024  & -0.004 & -0.004 & -0.011 & -0.011  \\
$L_{1t}, L_{2t}$  & 0.121 & 0.100 & 0.021& 0.150 & 0.000 & -0.004 & 0.006 & 0.001\\
\cline{2-9}
& \multicolumn{8}{c}{$N_1=25, N_2=75,T=100$} \\
\cline{2-9}
$G_t$ & 0.023 & 0.023 & 0.000& 0.020  & 0.051 & 0.054 & 0.019 & 0.022\\
$L_{1t}$ & 6.080 & 5.151 & 0.929 & 6.528 & 2.509 & 2.759 & 2.358 & 2.608\\
$L_{2t}$  & 1.139 & 1.104 & 0.035& 0.983 & 0.715 & 0.774 & 0.786 & 0.844\\
$G_t, L_{1t}$ & -0.003 & -0.008 & 0.005& 0.008  & 0.007 & 0.008 & -0.003 & -0.002 \\
$G_t, L_{2t}$ & -0.027 & -0.030 & 0.002 & -0.024 & -0.005 & -0.004 & -0.023 & -0.023  \\
$L_{1t}, L_{2t}$  & 0.156 & 0.135 & 0.020 & 0.150& 0.000 & -0.004 & 0.020 & 0.017\\
\cline{2-9}
& \multicolumn{8}{c}{$N_1=300, N_2=300,T=500$} \\
\cline{2-9}
$G_t$ & 0.003 & 0.003 & 0.000 & 0.003 & 0.009 & 0.009 & 0.004 & 0.004\\
$L_{1t}$ & 0.260 & 0.258 & 0.002 & 0.254 & 0.235 & 0.248 & 0.246 & 0.259\\
$L_{2t}$  & 0.274 & 0.272 & 0.002& 0.266 & 0.269 & 0.283 & 0.256 & 0.270\\
$G_t, L_{1t}$  & -0.002 & -0.002 & 0.000& -0.002 & 0.001 & 0.001 & -0.002 & -0.002 \\
$G_t, L_{2t}$  & -0.002 & -0.002 & 0.000& -0.002 & 0.000 & 0.000 & -0.002 & -0.002  \\
$L_{1t}, L_{2t}$  & -0.018 & -0.017 & 0.000 & -0.018& 0.000 & 0.000 & -0.018 & -0.018\\
\hline
\end{tabular}
\caption{The table reports the averages of empirical MSE $(\overline{\mathbf{MSE}}^{(emp)})$, covariances, and cross-product biases matrices (first three columns), together with the finite-sample approximations of the elements of the asymptotic MSE matrix computed using the true model parameters and estimated using HR without and with subsampling correction (HRS), and FPR without and with subsampling correction (FPRS). The DGP corresponds to an ML-DFM with one global factor and one group-specific factor for each of the two groups of variables, and cross-sectionally heteroscedastic and correlated idiosyncratic components. The factors are extracted using SLS. All entries are multiplied by 10.} 
\label{tab:MC_3}
\end{table}

\begin{figure}[h!]
\begin{center}
\begin{subfigure}[b]{1\linewidth}
\includegraphics[trim=15mm 15mm 0mm 18mm, clip, width=1\textwidth]{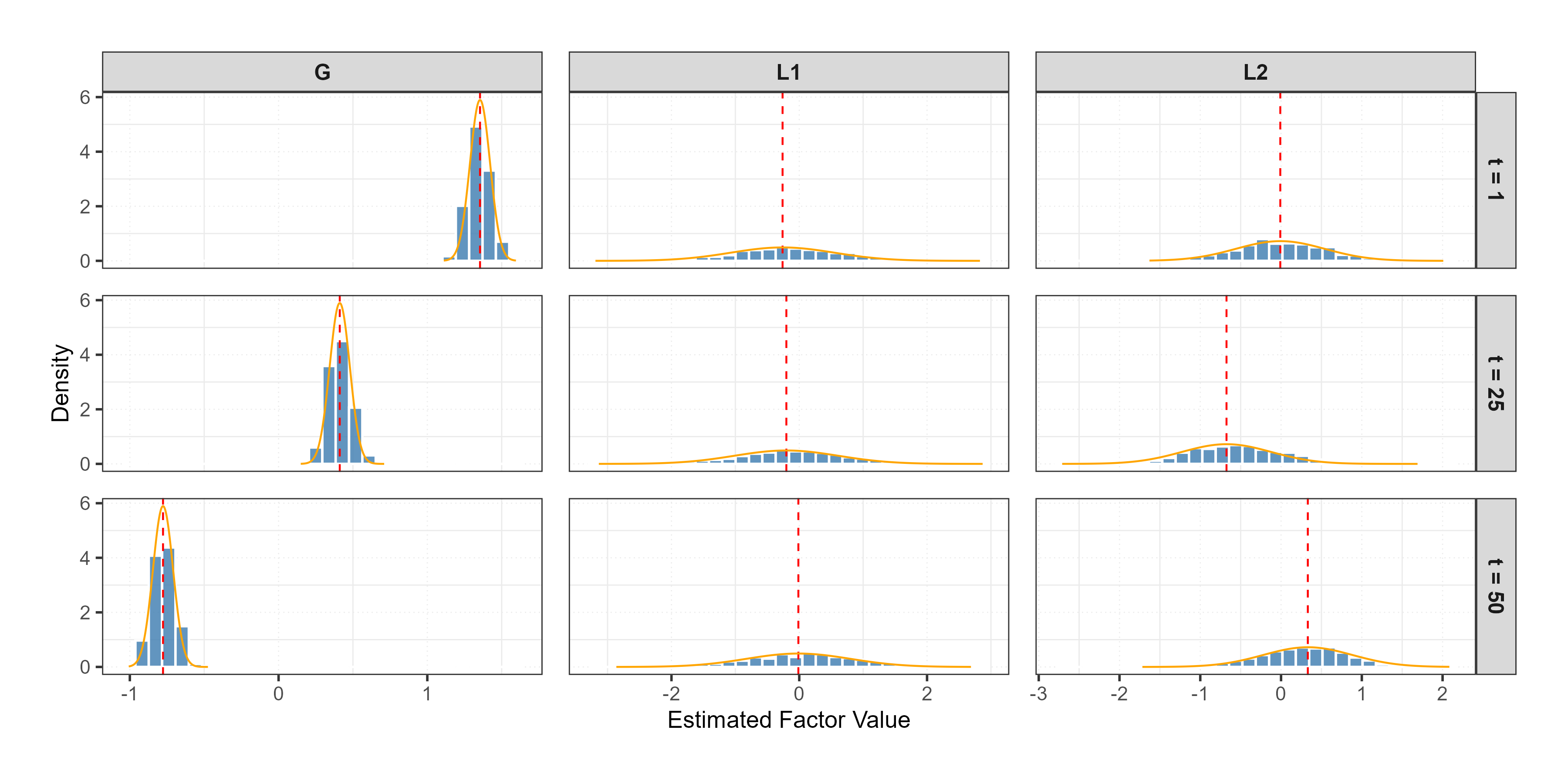}
\caption{$N_1=N_2=25, T=50$}
\end{subfigure}
\begin{subfigure}[b]{1\linewidth}
\includegraphics[trim=15mm 15mm 0mm 18mm, clip, width=1\textwidth]{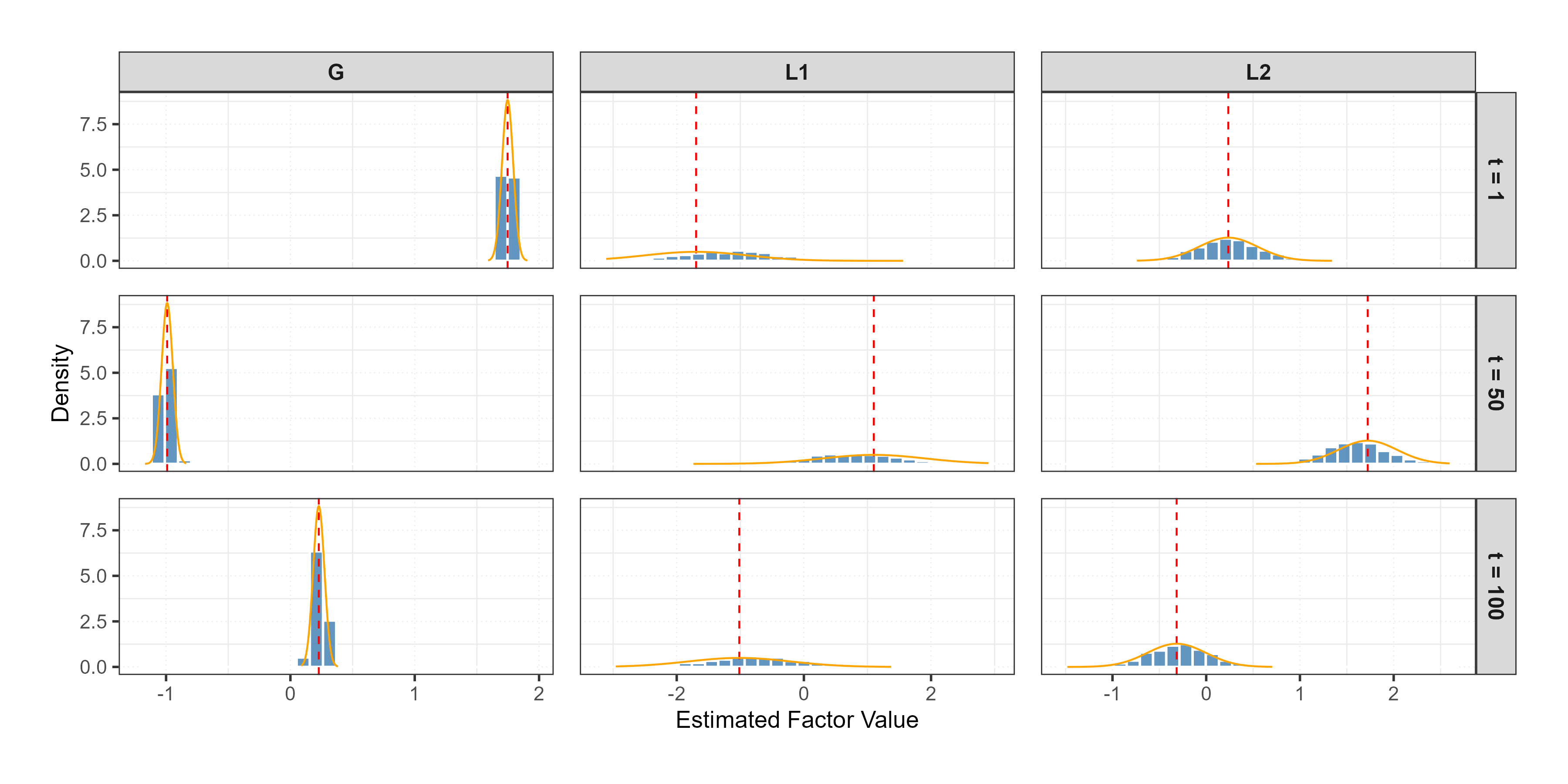}
\caption{$N_1=25, N_2=75, T=100$}
\end{subfigure}
\begin{subfigure}[b]{1\linewidth}
\includegraphics[trim=15mm 15mm 0mm 18mm, clip, width=1\textwidth]{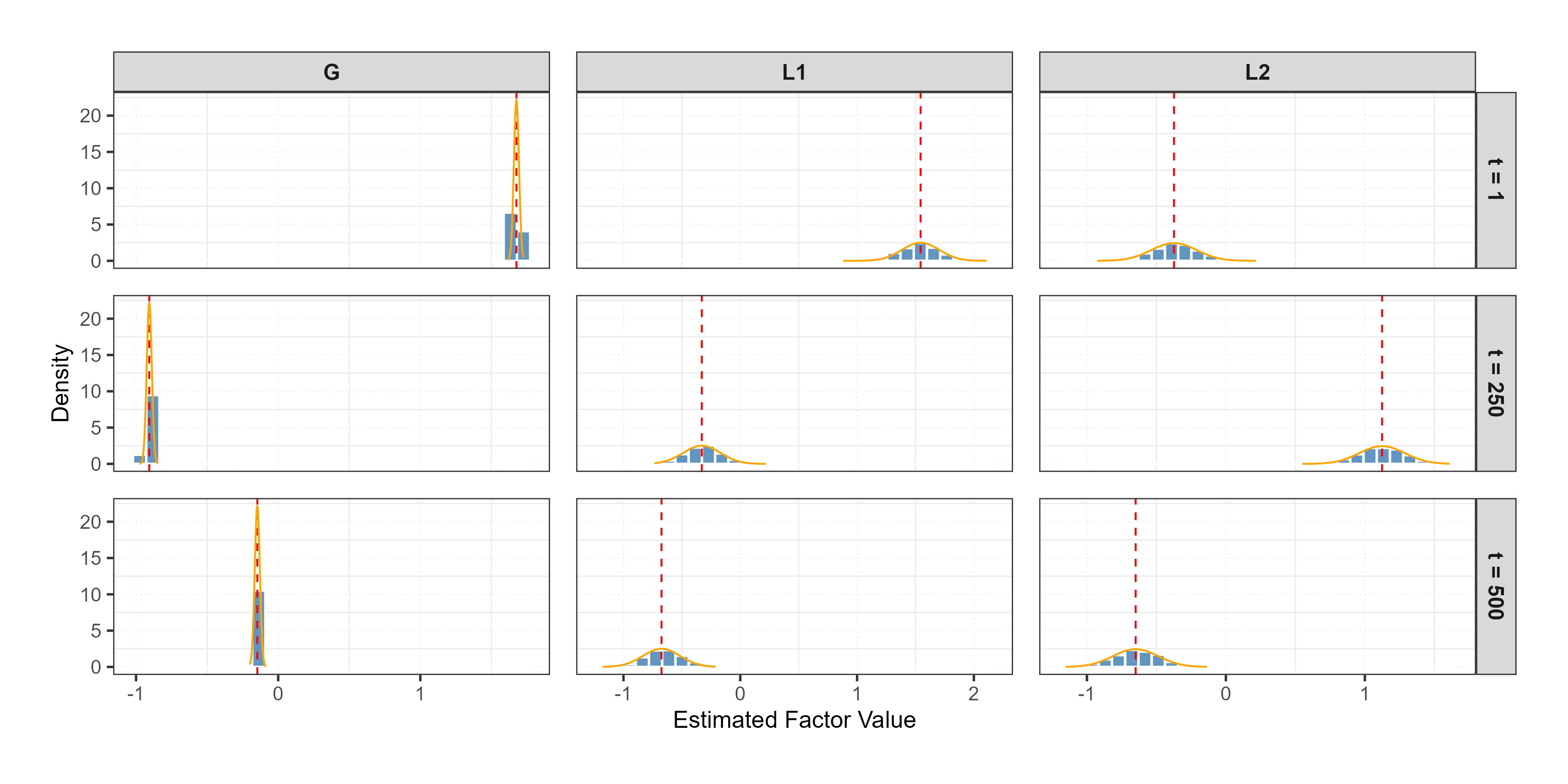}
\caption{$N_1=N_2=300, T=500$}
\end{subfigure}
\caption{Histograms of the estimated global factor, $G_t$ (first column), specific factor of the first block $L_{1t}$ (second column), and specific factor of the second block, $L_{2t}$ (third column), at three particular moments of time: i) $t=1$ (first column); ii) $t=T/2$ (second column); and iii) $t=T$ (third column). The red vertical line represents the value of the true factor. The idiosyncratic errors are cross-sectionally correlated and heteroscedastic.}
\label{fig:densities_3}
\end{center}
\end{figure}

\section{Final considerations}
\label{section:final}

The results in this paper are relevant for factor inference in ML-DFMs when the factors are estimated using SLS, which should be based on the asymptotic distribution of the SLS estimator. This distribution has been conjectured to coincide with the asymptotic distribution derived by Bai (2003) for the PC estimator in DFMs. Using Monte Carlo simulations across a wide range of empirically relevant temporal and cross-sectional dimensions and under several idiosyncratic covariance structures, we provide evidence supporting this conjecture. As far as we are concerned, this is the first paper using simulation to analyse not only the finite sample properties of point SLS estimates of the factors but also their distribution.

As a riveting addition contribution, we assess the performance of several estimators of the asymptotic MSE matrix proposed in the literature and show that the FPR estimator performs best when the idiosyncratic components are cross-sectionally correlated, without showing deterioration when they are not. We also illustrate the relevance of applying subsampling corrections to account for the uncertainty in estimating factor loadings.

The extension of the results in this paper to ML-DFMs with overlapping factors is in our research agenda.

\section*{Acknowledgements}

All authors acknowledge financial support from the Spanish National Research Agency (Ministry of Science and Technology) Project PID2022-139614NB-C22/AIE/10.13039/501100011033 (MINECO/FEDER). Any remaining errors are, of course, our responsibility.

\section*{Appendix A: Identification of ML-DFMs}

Consider the ML-DFM in (\ref{eq:ML-DFM}) with $S=2$ and $r_g=r_1=r_2=1$. The matrix of loadings is given by
\begin{equation*}
\mathbf{\Lambda}^{\dagger}=\begin{pmatrix}
    \lambda^{\dagger}_{11} & \lambda^{\dagger}_{21} & 0\\
    \lambda^{\dagger}_{12} & \lambda^{\dagger}_{22} & 0\\
    \dots & \dots & \dots\\
    \lambda^{\dagger}_{1N_1} & \lambda^{\dagger}_{2N_1} &0\\
    \lambda^{\dagger}_{1N_1+1} & 0 & \lambda^{\dagger}_{3N_1+1}\\
    \lambda^{\dagger}_{1N_1+2} & 0 & \lambda^{\dagger}_{3N_1+2}\\
    \dots & \dots & \dots\\
    \lambda^{\dagger}_{1N} & 0 & \lambda^{\dagger}_{3N}\\
\end{pmatrix}
\end{equation*}

Consider the following rotation
$$\mathbf{\Lambda}^*=\mathbf{\Lambda}^{\dagger} \mathbf{H}$$
If $\mathbf{\Lambda}^*$ satisfies the zero restrictions of the ML-DFM, then the $3 \times 3$ rotation matrix $\mathbf{H}$ should be such that
\begin{equation*}
    \mathbf{H}=\left[\begin{matrix}
    h_{11} & 0 & 0\\
    h_{21} & h_{22} & 0 \\
    h_{31} & 0 & h_{33}
    \end{matrix}  \right],
\end{equation*}
which implies $r_g^2+ r_1^2 + r_2^2 + r_g\left(r_1+r_2 \right)=5$ restrictions. These restrictions are: (i) the unitary modulus of each of the factors (3 restrictions); (ii) the orthogonality of the global factor and each of the two group-specific factors (2 restrictions). 

Consider now the ML-DFM in (\ref{eq:ML-DFM}) with $S=2$ and $r_g=2$, $r_1=2$, and $r_2=1$. The matrix of loadings is given by
\begin{equation*}
\mathbf{\Lambda}^{\dagger}=\begin{pmatrix}
    \lambda^{\dagger}_{11} & \lambda^{\dagger}_{21} & \lambda^{\dagger}_{31} & \lambda^{\dagger}_{41} & 0\\
    \lambda^{\dagger}_{12} & \lambda^{\dagger}_{22} & \lambda^{\dagger}_{32} & \lambda^{\dagger}_{42} & 0\\
    \dots & \dots & \dots & \dots & \dots & \\
    \lambda^{\dagger}_{1N_1} & \lambda^{\dagger}_{2N_1} & \lambda^{\dagger}_{3N_1} & \lambda^{\dagger}_{4N_1} & 0\\
    \lambda^{\dagger}_{1N_1+1} & \lambda^{\dagger}_2N_1 & 0 & 0 & \lambda^{\dagger}_{5N_1+1}\\
    \lambda^{\dagger}_{1N_1+2} & \lambda^{\dagger}_{2N_1+2} & 0 & 0 & \lambda^{\dagger}_{5N_1+2}\\
    \dots & \dots & \dots & \dots & \dots\\
    \lambda^{\dagger}_{1N} & \lambda^{\dagger}_{2N} & 0 & 0 & \lambda^{\dagger}_{5N}\\
\end{pmatrix}
\end{equation*}

Consider the following rotation
$$\mathbf{\Lambda}^*=\mathbf{\Lambda}^{\dagger} \mathbf{H}$$
If $\mathbf{\Lambda}^*$ satisfies the zero restrictions of the ML-DFM, then the $5 \times 5$ rotation matrix $\mathbf{H}$ should be such that
\begin{equation*}
    \mathbf{H}=\left[\begin{matrix}
    h_{11} & h_{12} & 0 & 0 &0 \\
    h_{21} & h_{22} & 0 & 0 & 0 \\
    h_{31} & h_{23} & h_{33} & h_{34} & 0\\
    h_{41} & h_{42} &  h_{43} & h_{35} &0\\
    h_{51} & h_{52} & 0 & 0 & h_{55}
    \end{matrix}  \right],
\end{equation*}
which implies $r_g^2+ r_1^2 + r_2^2 + r_g\left(r_1+r_2 \right)=15$ restrictions. These restrictions are: (i) the two global factors are orthonormal and their corresponding loadings are orthogonal (4 restrictions); (ii) the two group-specific factors of the first group are orthonormal and their corresponding loadings are orthogonal (4 restrictions); (iii) the unitary modulus of the group-specific factor of the second group (1 restriction); (iv) the orthogonality of each of the two global factors and each of the three group-specific factors (6 restrictions).


\end{document}